\documentclass[11pt]{paper}

\usepackage{graphicx}
\usepackage{multirow}
\usepackage{url}

\usepackage{color}

\usepackage{geometry}
\newcommand{\mytablesize}{\scriptsize}

\begin{document}

\title{\center{Social Interactions or Business Transactions?\\
\Large{What customer reviews disclose about Airbnb marketplace}}}

\author{
	Giovanni Quattrone, 
	\small{\em Middlesex University (UK), University of Turin (Italy), \url{g.quattrone@mdx.ac.uk}}\\
	Antonino Nocera,
	\small{\em University of Pavia (Italy), \url{antonino.nocera@unipv.it}}\\
	Licia Capra,
	\small{\em University College London (UK), \url{l.capra@ucl.ac.uk}}\\
	Daniele Quercia,
	\small{\em King's College (UK), Nokia Bell Labs (UK), \url{daniele.quercia@kcl.ac.uk}}
}

\maketitle

\begin{abstract}
Airbnb is one of the most successful examples of sharing economy marketplaces. With rapid and global market penetration, understanding its attractiveness and evolving growth opportunities is key to plan business decision making. There is an ongoing debate, for example, about whether Airbnb is a hospitality service that fosters social exchanges between hosts and guests, as the sharing economy manifesto originally stated, or whether it is (or is evolving into being) a purely business transaction platform, the way hotels have traditionally operated. To answer these questions, we propose a novel market analysis approach that exploits customers' reviews. Key to the approach is a method that combines thematic analysis and machine learning to inductively develop a custom dictionary for guests' reviews. Based on this dictionary, we then use quantitative linguistic analysis on a corpus of 3.2 million reviews collected in 6 different cities, and illustrate how to answer a variety of market research questions, at fine levels of temporal, thematic, user and spatial granularity, such as {\em (i)} how the business vs social dichotomy is evolving over the years, {\em (ii)} what exact words within such top-level categories are evolving, {\em (iii)} whether such trends vary across different user segments and {\em (iv)} in different neighbourhoods.
\end{abstract}

\section{Introduction}

The sharing economy, also known as peer-to-peer or collaborative economy, is an economic model based on a distributed network of individuals, directly accessing each other underused assets. Airbnb is one of the most successful examples of such model, with hosts renting out their unused rooms or entire properties by directly engaging in computer-mediated transactions with potential guests. Since its creation in 2008, Airbnb has been experiencing exponential growth, which continues to date. According to recent statistics,\footnote{https://ipropertymanagement.com/airbnb-statistics} the company is currently operating in more than 65,000 cities worldwide, with over 6M listings to choose from, and serving over 2M  people on any given night. Airbnb marketplace is not only growing but also very rapidly evolving: for example, while millenials still make up the largest portion of user share at 60\%, in the last two years the fastest growing host demographic has been in senior hosts over 60, with a growth rate of 102\%. On top of demographic diversification, Airbnb has been experiencing geographic habit diversification, too: for example, the average Berlin guest stays for 6.3 nights, as opposed to the average Amsterdam guest who stays for 3.9 nights only.

One of the challenges that companies like Airbnb face is to understand their attractiveness, as well as their evolving market opportunities, in the face of such rapid and diversifying growth rates. Traditional market research techniques, based on customer surveys and focus groups, offer very detailed insights that can help inform business decision making, but require substantial financial and time investments. As a result, their use in the sharing economy context is limited, due to the fast-evolving and global nature of most such markets. In this setting, more agile techniques are needed to allow companies to strategise promptly. For example, there is an ongoing debate about whether Airbnb is a hospitality service that fosters social exchanges between hosts and guests, as the sharing economy manifesto\footnote{http://www.thepeoplewhoshare.com/} originally stated, or whether it is (or is evolving into being) a purely business transaction platform, the way hotels have traditionally operated. Being able to assess to what extent Airbnb customers value social interactions \emph{vs.} business transactions has important implications for how the company may decide to operate, and compete, in the hospitality service. Given Airbnb different usage patterns in different cities, such market analysis needs to be performed separately in each geographic context the company operates; furthermore,  because of the rapidly evolving demographics of its customers, the analysis needs to be repeated frequently, to  capture varying trends.

In this paper, we propose a scalable market analysis approach, to complement and enrich traditional ones. Instead of collecting primary data via interviews, focus groups and surveys, our approach exploits ready-available secondary data that most sharing economy platforms like Airbnb possess: a continuous stream of reviews that peers leave upon completion of a service exchange. Key to our approach is a new semi-supervised method to inductively develop platform-specific dictionaries starting from peers' reviews. The method combines qualitative thematic analysis with quantitative machine learning techniques in a novel way, and enables the construction of a dictionary that captures topics disclosed in customers' reviews at different levels of granularity. Based on this purpose-built dictionary, we then define robust topic-adoption metrics that enable us to explore a variety of market research questions, at fine levels of thematic, temporal and spatial granularity.

%Unlike unsupervised topic detection techniques (e.g., Latent Dirichlet Allocation model~\cite{jelodar2019latent}), our approach does not suffer from the problem of over-fitting that is common when text length is short, as it is often the case in reviews (e.g., \cite{cuong2019eliminating}). Furthermore, unlike approaches that rely on general-purpose dictionaries such as LIWC~\cite{pennebaker2001linguistic}, our approach affords the exploration of platform-specific market research questions (rather than platform-agnostic explorations about, for example, sentiment analysis and mood detection~\cite{tausczik2010psychological}). 
%\notegq{The intro was too long. I removed the sentence ``Unlike unsupervised topic detection ...''. Pls check if you agree}
%\notelc{ok but we need to move it elsewhere}
 
We specifically illustrate our proposed market analysis approach using the case of Airbnb, and while doing so, we make the following two main contributions:

\paragraph{(1) Dictionary construction.} We gather 3.2M Airbnb guest reviews about 176K distinct listings, spread across 6 different cities (London, Manchester, New York, San Francisco, Melbourne, Sydney),  written between 2010 and 2019 (Section~\ref{sec:data}). These cities have been chosen so to span different continents (America, Europe, Oceania), later affording us the ability to explore whether trends are geographically bounded or not. Note that, at this stage, we are focusing on reviews written in English only; these represent 90\% of all reviews left for properties in these cities. We then analyse these reviews using a combination of thematic analysis and machine learning, and build a dictionary that is capable of classifying words (unigrams) at three levels of granularity: two top-level categories (i.e., `social' interactions \emph{vs.} `business' transactions), four distinct sub-categories and 13 subsub-categories (Section~\ref{sec:dict}). %In this instantiation of the method, we did not make use of the reviews that hosts leave about guests, since these would not contribute much knowledge to the market research investigation we are currently pursuing. \notedq{not sure how to make this last sentence more concrete}\notelc{suggest cutting completely}

\paragraph{(2) Market analysis.} We illustrate how to use the purpose-built dictionary, in combination with robust topic-adoption metrics, to understand to what extent Airbnb guests discuss the social aspect \emph{vs.}  the business aspect of their hospitality experience (Section~\ref{sec:linguistic}). We do this by exploring four different market research questions that illustrate the ability of our dictionary and analytical approach to address questions at varying levels of detail, while also scaling easily over time and geographic location. We find that, across the 6 cities analysed, business aspects are increasingly being discussed in guests' reviews, while social aspects are steadily declining (Section~\ref{sec:years}). This trend is happening not just at the top-level categories (business \emph{vs.} social), but across all words in our lexicon (Section~\ref{sec:words}). We then segment Airbnb hosts according to the time they joined the platform, and discover that those who joined at the very beginning (i.e., the so called `innovators'~\cite{rogers2010diffusion}), are those receiving guests' reviews that most dwell on the social aspects of their hospitality experience, and they remain so over the years. On the contrary, hosts who joined the platform later (`early adopters' and `early majority'), consistently receive more business-dominated reviews across all cities (Section~\ref{sec:segments}). Finally, we zoom in within each city, to understand whether there is market diversification in different neighbourhoods, and discover that properties in areas of low Airbnb penetration (less tourist areas) receive reviews that discuss social aspects of the experience significantly more than those in areas with higher Airbnb penetration (more tourist areas). Once again, this pattern is consistent in all cities analysed, despite them being located in different countries/continents (Section~\ref{sec:neighborhoods}). 

We conclude this paper with a discussion about practical uses of the proposed method, its current limitations, and possible future developments (Section~\ref{sec:conclusion}).

\section{Related Work}

%\notegq{Change and shorten it. We can start saying that Airbnb has been studied a lot in the recent years from the impact to the society (cite!), to the motivation behind the sharing economy platform's uptake (cite!) to the rating and reviews. We need to expand and discuss in detail only this section. That is, scholars have studied sentiment, review scores, initial linguistic analysis and show what they have found. No linguistic studies on the reviews have been proposed to perform market analysis. Here it is where our work is located.}

Sharing economy platforms like Airbnb have been extensively studied in the past, following two broad lines of inquiries.

A first line of inquiry has analysed the relationship between {\em sharing economy services and society}, specifically at the level of  cities~\cite{Zervas2016rise,zervas2015impact,fang2016effect,quattrone2016airbnb,liu2018impact,roma2019sharing,heo2019happening}. Several studies have looked into the relationship between these novel services and their traditional counterparts, with findings that often varied depending on geographic location: some scholars found that these new services only marginally disrupt their established counterparts (e.g., Uber \emph{vs.} taxis, Airbnb \emph{vs.} hotels)~\cite{Zervas2016rise}. As an example, in London, the geographical overlap between Airbnb properties and hotels was found to be marginal~\cite{quattrone2016airbnb}; furthermore, sharing economy services were found to bring positive effects to the broad tourism industry~\cite{fang2016effect}. Other scholars found opposite results instead: a study performed in Budapest showed that Airbnb and hotels were located in the same central areas,  causing fierce competition between the two~\cite{boros2018airbnb}. Other studies have looked at the relationship between Airbnb and the housing/rental market~\cite{wachsmuth2018airbnb,shabrina2019airbnb}, with findings suggesting that Airbnb is accelerating an ongoing processes of gentrification in London.

%In a first line of enquiry, scholars have analysed the impact of sharing economy services on our cities~\cite{Zervas2016rise,zervas2015impact,fang2016effect,quattrone2016airbnb,liu2018impact,zervas2015impact,roma2019sharing,heo2019happening}. Findings indicate that, these new services only marginally disrupt their established counterparts (e.g., Uber vs. taxis, Airbnb vs hotels)~\cite{Zervas2016rise}; as an example, research shows that in London the geographical overlap between Airbnb properties and hotels is only marginal~\cite{quattrone2016airbnb}. Furthermore, sharing economy services are able to bring a number of positive effects to the broad tourism industry~\cite{fang2016effect}. Other scholars have criticised Airbnb instead, as a possible cause of gentrification~\cite{wachsmuth2018airbnb,shabrina2019airbnb}. Findings show that in London Airbnb is associated with areas that have a high proportion of privately rented properties; such a phenomenon can reach up to 20\% in some neighbourhoods, further exacerbating the process of gentrification~\cite{shabrina2019airbnb}.

A complementary line of inquiry has focused on the relationship between {\em sharing economy services and people}. Several studies have looked into motivational factors for user participation in such platforms. Using online surveys and host/guest interviews, these investigations have revealed that financial benefits are an important factor for Airbnb hosts to join such platforms, but they do not represent the only factor, as business (financial) reasons and social reasons are intertwined with one other~\cite{satama2014consumer,ikkala2015monetizing,lampinen2016hosting,guttentag2018tourists,bancoro2018tourists,lin2019spend}. Whether this is changing over time, and in different locations, is hard to answer, since the primary data used to perform such studies (e.g., survey data) is very costly to obtain (both financially and in terms of time). Other studies have used ready available data from within these online platforms instead, primarily to study user satisfaction with the service provided. An analysis of Airbnb ratings has revealed that 95\% of properties in Airbnb boast an average user-generated rating above 4.5 stars~\cite{bridges2018if,zervas2015first}; this is in sharp contrast with platforms like TripAdvisor, where the average star rating is 3.8~\cite{zervas2015first}. Sentiment analysis conducted on reviews seemed to corroborate this finding~\cite{fradkin2015bias,alsudais2019large,lawani2018reviews,luo2018airbnb}, although the authors caution against a phenomenon of ``socially induced reciprocity'' which may occur when peers interact socially with one another, leading to negative information being omitted from reviews. Scholars have used sentiment analysis on user reviews to shed light on price dynamics too, revealing that the price of Airbnb properties is greatly influenced by their review score, after controlling for characteristics of the room and features of the neighbourhood~\cite{lawani2018reviews}.

Recently, reviews have increasingly been used as main data source in sharing economy platform studies~\cite{luo2019understanding,joseph2019analyzing,martin2018modelling,cheng2019airbnb,lee2019analysing}, not only because they are ready available, but also because, with over 70\% of guests writing a review after a stay~\cite{fradkin2017determinants}, they can offer very good coverage of peers' experiences in such platforms. For example, in~\cite{jung2016social} researchers collected a sample of hosts' profiles and guests' reviews in AirBnB and Couchsurfing; after manually labelling and analysing them, they found initial evidence that the primary shared asset in AirBnB is the house (i.e., its facilities, location, neighbourhood), while in Couchsurfing it is the human relationship (i.e., host-guest interaction, experience, self-description, motivation). This finding is corroborated by another study that used interviews as primary data source instead: in~\cite{klein2017quality}, 17 users who had participated in both Airbnb and Couchsurfing were interviewed, revealing that Airbnb peers require higher quality services, and put more emphasis on places over people. The same study~\cite{klein2017quality} also analysed 5k random reviews from Couchsurfing and Airbnb using the general-purpose LIWC dictionary~\cite{pennebaker2001linguistic}. Once again, results confirmed that Airbnb reviews are more business oriented, whereas Couchsurfing reviews are more person-oriented; since the LIWC dictionary is platform-independent, it is not possible to delve deeper into this business \emph{vs.} social dichotomy. To zoom in further, recent studies have taken an orthogonal approach, mining reviews in an unsupervised fashion, and analysing platform-specific emerging topics: for example,  in~\cite{cheng2019airbnb} topics such as `location', `amenities' and `host' appear to automatically emerge; in~\cite{luo2019understanding}, the five most common aspects of Airbnb reviews that emerge seem to be the communication between guest and host, the experience of the rental, the location of the property, the service offered, and the value of the property. Both studies suggest once again that the nature of Airbnb is mainly about accessing assets rather than sharing them.

In this paper, we further expand on this latter line of inquiry, and propose a mixed-method approach that combines thematic analysis of guest reviews with unsupervised machine learning techniques, to inductively build a dictionary that enables fine-grained and scalable market analysis of platforms such as Airbnb. 
%Unlike approaches based on general purpose dictionaries like LIWC~\cite{pennebaker2001linguistic}, our approach supports platform-specific investigations; also, unlike completely unsupervised approaches, ours offer a hierarchical taxonomy that supports investigations at different levels of detail. 
Unlike unsupervised topic detection techniques (e.g., Latent Dirichlet Allocation model~\cite{jelodar2019latent}), our approach does not suffer from the problem of over-fitting that is common when text length is short, as it is often the case in reviews (e.g., \cite{cuong2019eliminating}). Furthermore, unlike approaches that rely on general-purpose dictionaries such as LIWC~\cite{pennebaker2001linguistic}, our approach affords the exploration of platform-specific market research questions (rather than platform-agnostic explorations about, for example, sentiment analysis and mood detection~\cite{tausczik2010psychological}).
Before presenting our proposed method, we briefly introduce the dataset we collected.

\section{Dataset}
\label{sec:data}

We gathered Airbnb data from the ``Inside Airbnb'' organisation (\url{http://insideairbnb.com/}), containing snapshots of Airbnb listings and reviews around the world collected at regular time intervals (typically, at least once per quarter from 2015, and  more often in the last couple of years). 

On June 3rd 2019, we gathered all the listings and reviews associated with six different cities: Greater Manchester (U.K.), London (U.K.), Melbourne (Australia), New York City (U.S.), San Francisco (U.S.), Sydney (Australia). We selected these cities for the following two reasons. First, we did not want to add the inherent noise incurred when performing language translation; we thus favoured cities in English-speaking countries, for which we expected the vast majority of reviews to be written in English. Second, within this constraint, we wanted to consider cities belonging to different countries and continents, so to later explore whether our findings are country/continent bounded or they  generalize.

We initially collected 3.9 million Airbnb guest reviews associated with 176 thousand distinct listings. To gain confidence in the validity of the data, we selected 10 random listings, along with their associated reviews in each city and verified their existence on the original Airbnb platform. After this preliminary check, we analysed review length distribution, and removed reviews that were either too short or too long (less than 5 words and more than 175 words -- which are about 8\% of the original reviews). We further removed reviews automatically generated by the system in case of a cancellation (around 2\%); reviews without a year and without comments (less than 1\%); reviews generated by power users (i.e., guests who wrote more than 10 reviews) who may bias results (less than 1\%), and finally non English reviews (around 5\% of reviews removed). We ended up with a dataset comprising 3.2 million guests' reviews, whose composition by city and by year is shown in Table~\ref{tab:ReviewsByCityYear}. 

\begin{table}[!t]
   \centering
   \mytablesize 
   \begin{tabular}{ cc }   
      \begin{tabular}{lr}
         \em City & \em \# Reviews \\
         \hline \hline
         Greater Manchester & 91,967\\
         London & 992,638\\
         Melbourne & 469,906\\
         New York City & 883,280\\
         San Francisco & 286,592\\
         Sydney & 438,491\\
         \hline \hline
         \end{tabular} &  
         \hspace{10mm}
         \begin{tabular}{lr}
         \em Year & \em \# Reviews \\
         \hline \hline
         2010 & 1,805\\
         2011 & 7,398\\
         2012 & 20,091\\
         ... \\
         2017 & 706,556\\
         2018 & 1,101,528\\
         2019 & 500,834\\
         \hline \hline
      \end{tabular} \\
   \end{tabular}
   \caption{Reviews by city and by year}
   \label{tab:ReviewsByCityYear}
\end{table}

\section{Dictionary construction, adoption and validation}
\label{sec:dict}

In this section, we present a mixed-method approach that combines thematic analysis with machine learning techniques to inductively build a platform-specific (in this case, Airbnb) dictionary that affords us the ability to group the lexicon used in Airbnb guest reviews into categories concerning `social interactions' \emph{vs.} `business transactions' at different levels of granularity  (Section~\ref{subsec:buildingDict}). We  then define metrics to be computed on top of this dictionary (Section~\ref{subsec:adoptionDict}), and report on dictionary and metric validation steps we have conducted (Section~\ref{subsec:validatingDict}).

%that will enable us to explore a variety of market research questions, at different level of thematic, temporal and spatial granularity. Before illustrating examples of such market investigations (Section~\ref{sec:linguistic}), we will report on a number of dictionary and metric validation steps we have conducted (Section~\ref{subsec:validatingDict}).

\subsection{Building a Dictionary}
\label{subsec:buildingDict}

We built our dictionary in five steps:  first, we developed a coding scheme by performing thematic analysis of a  random sample of 100 Airbnb reviews (step~1); second, we refined and validated the coding scheme by means of a crowd-sourcing study conducted on the Crowdflower\footnote{Crowdflower is a crowd-sourced market of online workforce to clean, label and enrich data: \url{https://www.crowdflower.com/}.} platform (step~2), where we asked crowd-workers to label another random set of 100 reviews. Third, we conducted a second study on Crowdflower, this time asking crowd-workers to label a larger set of 1,500 reviews, using the identified themes (step~3). Using natural language processing techniques, we then defined a lexicon of the words most representative of each such theme (step~4). Finally (step~5), using hierarchical clustering techniques, we grouped together these words into 13 distinct clusters, which represent a finer-grained refinement of the themes manually identified at steps~1 and~2. Our final dictionary comprises two level-1 categories (i.e., business vs social), refined into four level-2 (sub)categories, further refined into thirteen level-3 (subsub)categories, which semantically group together a lexicon of 355 words. We discuss the details of each step next.

\paragraph{Step~1. Developing a Coding Scheme.}

Using stratified sampling to cover all study years and cities, we sampled 100 Airbnb reviews. We broke down each review into its constituting sentences, and performed a thematic analysis over these. In a way similar to~\cite{braun2006using}, two independent annotators coded these resulting sentences by performing three steps: {\em (i)}~familiarising with the data,  {\em (ii)}~generating the initial codes and searching for themes among codes, and {\em (iii)}~defining themes. After a first round of coding, the two coders compared their results, and agreed on which themes to maintain, remove, amend, or merge. As a result, they agreed on five main themes named `property', `location', `business conduct', `personality', and `social interaction'. The first three are refinements of the theme `business' and the last two of the theme `social'.

%The theme `property' covered topics related to the property/room such as its cleanness, its decoration, size, or furniture. The theme `location' covered topics related to the neighbourhood where the property is situated such as nearby viewpoints, restaurants, metro stations, and so forth. The theme `professional conduct' captured topics related to the professional conduct of host such as how good their communication has been, or whether the host has been flexible with check-in/check-out times. The theme `personality' referred to the personality of the host, and finally the theme `social interaction' referred to guest/host spending social time together. Intuitively, the first three themes (i.e., `property', `location', `business conduct') refer to business-oriented discussion topics, while the latter two (i.e., `personality', `social interaction') refer to social-oriented ones.

\paragraph{Step~2. Validating the Coding Scheme.}

To gain confidence in the validity of the coding scheme, we asked crowd-workers to annotate sentences extracted from a new sample of 100 Airbnb reviews using these five themes. In particular, we prepared a  Crowdflower page that consisted of three sections: {\em (i)}~a list that showed our five themes; {\em (ii)}~for each theme, actual examples of Airbnb reviews manually labelled by us; and  {\em (iii)}~new Airbnb sentences to be labelled. We paid 0.01\$ per annotation, and each Airbnb sentence was independently annotated by at least four different workers. We computed the Fleiss' kappa agreement score for the five themes~\cite{fleiss1971measuring}, and two of them (i.e., `personality' and `social interaction') had a Fleiss' kappa score less than 0.5. We merged these two themes into one, resulting in four themes: `property', `location', `professional conduct' and `social interaction'. To ascertain the effectiveness of coding with those four themes, we again asked crowd-workers to annotate a new sample of sentences extracted from yet another 100 Airbnb reviews. All four themes resulted in a Fleiss' kappa score higher than 0.5, suggesting their validity.

\paragraph{Step~3. Labelling Reviews.} 

We were then ready to label a larger set of Airbnb reviews using the identified four themes. We used again Crowdflower to annotate unlabelled sentences extracted from a new set of 1,500 reviews. We gathered 22,975 distinct annotations of 4,062 sentences. We kept those sentences on which at least 75\% of annotators agreed -- so to have high confidence that the words inferred from these sentences are reliable -- and ended up with a set of 1,868 sentences having high agreement. The second column of Table~\ref{tb:StatsCodingScheme} shows the frequency of occurrence of each of the four themes in these sentences. The most popular theme was `property', followed by `location' and `professional conduct'; `social interaction' was the least frequent theme instead. 

\begin{table}[!bht]
\tabcolsep 4pt
\mytablesize
\centering
\begin{tabular}{ l c | c c }
   \hline \hline
   \em Theme & 
   \em  Frequency &
   \em Initial words & 
   \em Expanded words  \\
   \hline \hline
   Property             & 35\% & ~63      & ~77\\
   Location             & 28\% & ~97      & 109\\
   Professional Conduct & 23\% & 107      & 119\\
   Social Interaction   & 14\% & ~61      & ~68\\
   \hline \hline   
\end{tabular}
\caption{Inferred four themes along with their frequency, number of words in each theme before and after enrichment}
\label{tb:StatsCodingScheme}
\end{table}

\paragraph{Step~4. Building the Dictionary.}

To build a dictionary, we needed to identify a lexicon (that is, list of words) that could represent the four themes above.  We did so in a data-driven fashion. First, for each theme $\tau$, we split the 1,868 annotated sentences into two sets: {\em (i)} $Set_{\tau}$, that is the set of sentences labelled with the theme $\tau$ by at least three quarter of workers; and {\em (ii)} $Set_{\bar{\tau}}$, that is the set of sentences labelled with the theme $\tau$ by at most one worker. Second, we extracted all words from $Set_{\tau}$ and $Set_{\bar{\tau}}$. For each word  $w$, we computed two measures: $tf(w,\tau)$ and $tf(w,\bar{\tau})$, respectively denoting the term frequency of $w$ in $Set_{\tau}$ and in $Set_{\bar{\tau}}$. Finally, we computed $tf_{gain}(w,\tau) = \frac{tf(w,\tau)}{tf(w,\bar{\tau})}$. 

For each theme $\tau$, we then associated all the words $w$ such that $tf(w,\tau) \geq tf_{min}$, $tf(w,\tau) \leq  tf_{max}$ and $tf_{gain}(w,t) \geq tf_{gain}$, with $tf_{min}, tf_{max} \in [0,1]$ and $tf_{gain} \in [1, +\infty)$. The first two thresholds, $tf_{min}$ and $tf_{max}$, allowed us to remove extremely unpopular and extremely popular words respectively. The use of the last threshold $tf_{gain}$ enabled us to associate to a theme $t$ only those words that were {\em comparatively more popular} in $Set_{\tau}$ than in $Set_{\bar{\tau}}$. Since there is no ground-truth about what a dictionary should look like, automated parameter tuning was not viable. Rather, different thresholds needed to be manually tested and validated. To this purpose, we followed a methodology resembling the Elbow criterion~\cite{ketchen1996application}. Specifically, we considered the following threshold values: $tf_{min} = \{ 0.001, 0.01, 0.05 \}$, $tf_{max} = \{ 0.15, 0.30, 0.60, 1 \}$, $tf_{gain} = \{ 1.5, 2, 3, 4, 6 \}$. We started with the the most restrictive combination of $tf_{min}$, $tf_{max}$ and $tf_{gain}$; that is, the combination of parameters generating the smallest dictionary. This combination was $tf_{min} = 0.05$, $tf_{max} = 0.15$ and $tf_{gain} = 6$.  We then changed each threshold value iteratively, with each iteration adding a new set of words to the dictionary. We manually validated this added set of words and measured the ratio of noise; that is, the ratio of words that according to our (human) judgement were incorrectly assigned to a particular category. We stopped our search for the best combination of parameters when this ratio was significantly higher than the one identified at the previous step. We ended up with the following manually tuned thresholds: $tf_{min} = 0.01$, $tf_{max} = 0.15$ and $tf_{gain} = 3$. The third column of Table~\ref{tb:StatsCodingScheme} summarises the number of words that each theme contained at this point. %\notelc{i replaced category with theme, since we only start having categories at the next step}

%we used a grid search with $tf_{min} = \{ 0.001, 0.01, 0.1 \}$, $tf_{max} = \{ 0.15, 0.30, 0.60, 1 \}$, $tf_{gain} = \{ 1.5, 2, 3, 4, 6 \}$. We started with the the less restrictive combination of $tf_{min}$, $tf_{max}$ and $tf_{gain}$; that is, the combination of parameters creating the biggest dictionary (i.e., having the highest number of words for each category). This combination was $tf_{min} = 0.001$, $tf_{max} = 1$ and $tf_{gain} = 1.5$. Then we manually validated this dictionary by evaluating whether each of its words has been correctly or incorrectly assigned to a given category. We ended up with a manually refined dictionary which we considered as benchmark. Then, for each other parameter configuration we computed the f-measure\footnote{We remind the reader that in classification the f-measure score is the harmonic mean between precision and recall~\cite{van1979information}.} according to our benchmark. Finally, we chose the combination of $tf_{min}$, $tf_{max}$ and $tf_{gain}$ maximising f-measure so to have a balanced accuracy/completeness trade-off. We ended up with the following manually tuned thresholds, $tf_{min} = 0.01$, $tf_{max} = 0.15$ and $tf_{gain} = 3$. The third column of Table~\ref{tb:StatsCodingScheme} summarises the number of words that each theme contained at this point. %\notelc{i replaced category with theme, since we only start having categories at the next step}

We then used a word embedding machine learning technique (i.e., word2vec~\cite{goldberg2014word2vec}) to further enrich our initial lexicon. We started by training the technique on the whole corpus of 3.2M reviews, and mapped each word into a vector having 50 dimensions. For each word already present in our lexicon, we then computed a list of similar words, that is, a list of words having a cosine similarity higher than a threshold $th_{cos}$. We included these words as part of the lexicon of our dictionary if they were not already present. In so doing, we enriched our dictionary with words that are not frequently used in the 1,868 labelled sentences, but still widely used in the whole corpus of reviews (and similar to those previously derived from our labelled corpus). We used a procedure similar to the one described above to manually tune $th_{cos}$. The threshold values considered during this step were $th_{cos} = \{ 0.6, 0.7, 0.8, 0.9 \}$; the manually tuned value chosen in the end was $th_{cos}=0.7$. The last column of Table~\ref{tb:StatsCodingScheme} shows the total number of words belonging to each of the four themes after this enrichment step.

\paragraph{Step~5. Identifying categories at different levels of granularity.}

A manual inspection of our expanded lexicon revealed that several sub-themes could be identified within the four main ones that we manually coded at steps~1 and~2. For example, under theme `social interaction', we identified both words that refer to {\em whom} the peers interacted with (e.g., husband, wife, daughter) as well as {\em how} (e.g., meals together, talking). In order to offer a more fine-grained taxonomic structure on top of our lexicon, we used a clustering algorithm. For each of the 4 themes in turn, we took all the words associated with them and placed them in a single cluster. We then iteratively increased the number of clusters until the `optimal' number of clusters was found. We chose k-means as clustering algorithm, with the Elbow method~\cite{ketchen1996application} applied to find the optimal number of clusters. We ended up with 13 clusters: three clusters were  refinements of the `property' theme, four clusters were refinements of the `professional conduct' theme, and a further five of the `social interaction' theme. The `location' theme was mapped to a single cluster, without further refinement.  

Table~\ref{tab:StatsFinalDictionary} provides an overview of the final dictionary we built. Themes were directly mapped into a 3-tier hierarchical structure consisting of two level~1 categories (that is, `business' and `social'), four level~2 categories (that is, `property', `location', `professional conduct', and `social interaction'), and thirteen level~3 categories (those automatically inferred by our clustering analysis).\footnote{Note that the name of the sub-theme was assigned by us after clustering.} An example of lexicon for each category is also provided (the top five words by inverse order of term frequency).
%, together with the intra-cluster similarity computed for each level~3 category (the higher the value, the more cohesive the words inside the cluster). 
The full dictionary is available for download at \url{https://figshare.com/s/991c8677e3e9ce013774}. 

\begin{table*}[th]
\tabcolsep 3pt
\mytablesize
\centering
\begin{tabular}{ l | l | l | c l  }
   \hline \hline
   \em Categories lev.~1 & 
   \em Categories lev.~2 &    
   \em Categories lev.~3 & 
   \em \#Words & 
   \em Top 5 words (by term frequency) \\
   \hline \hline
   \multirow{9}{*}{Business} &              & Property type & ~17 & apartment, house, home, flat, private \\
   & Property    & Interiors & ~43 & clean, comfortable, room, bed, kitchen \\
   &             & Facilities & ~17 & water, hot, wifi, towels, tv \\
   
   \cline{2-5}

   & Location    & Location   & 109 & quiet, area, walk, located, restaurants  \\

   \cline{2-5}
   
   &                   & Communication & ~33 & communication, questions, quick, communicative, responded \\
   & Professional      & Logistics & ~22 & check, provided, arrival, late, keys \\
   &     Conduct       & Advice & ~11 & information, recommendations, tips, advice, suggestions \\
   &                   & Hospitality & ~35 & helpful, welcoming, available, accommodating, responsive \\

   \hline
   
         &              & People & ~24 & family, friend, husband, wife, daughter \\
         & Social       & Personality & ~22 & friendly, kind, warm, charming, sweet \\
 Social  & Interaction  & Sharing & ~~6 & share, sharing, experiences, stories, interests \\
         &              & Talking & ~~8 & chat, conversation, talking, chatting, moments \\
         &              & Meals & ~~8 & breakfast, delicious, fresh, dinner, meals \\
   \hline \hline   
\end{tabular}
\caption{Summary statistics of the final dictionary}
\label{tab:StatsFinalDictionary}
\end{table*}

Quite interestingly, the clusters corresponding to {\em property} directly matched the property description fields of Airbnb listings -- that is, property type (e.g., whether a house or a flat), internal layout (e.g., kitchen, bed, cozy), and facilities (e.g., wifi, tv, fridge). In terms of {\em professional conduct}, distinct elements have been detected: basic communication (e.g., questions, quick, responded), handling of logistics (e.g., check in, arrival), and provision of advice (e.g., tips, directions). 
%Together with {\em location}, these categories are indicative of reviews that `talk business' (i.e., {\em business category}). 
For {\em social interaction}, five level-3 categories have emerged from clustering, these being `people' (e.g., with whom the guests interact -- e.g., husband, wife), what their `personality' is (e.g.,  friendly, kind, warm), if/what they are `sharing' (e.g., share, stories, experiences), and the how -- `talking' (e.g., chat, talking, conversation) over a `meal' (e.g., breakfast, dinner together).
% This theme, with its sub-themes,  is indicative of reviews that `talk social' (i.e., {\em social category}).

\subsection{Adopting our dictionary}
\label{subsec:adoptionDict}

Having built the dictionary above, our next step is to define metrics operating with its categories (from level~1 to level~3), and its lexicon.

\paragraph{Metric operating on the dictionary categories.}

The first metric we define works at the category level, and it is called {\em adoption}. As its name suggests, it measures the adoption of a specific category on a given set of reviews. Specifically, let $R$ be a set of reviews (e.g., reviews left in a given year and/or city), let $r \in R$ be a specific review belonging to $R$, and let $c$ be the category (of any level, from level~1 to level~3) under consideration. Let us define as $W$ the set of words contained in $R$ and as $C$ the set of words belonging to category $c$. For each word $w \in W$ contained in the review $r$, we compute the logarithmically scaled term frequency $tf(w,r)$. For each pair $\langle w,r \rangle$, we define the percentage of adoption of a category $c$ associated with the review $r$ as:
   \begin{equation}
	\% adp(c, r) = \frac{ \sum_{w \in C}{tf(w,r)} }{ \sum_{w \in W}{tf(w,r)} } \times 100
   \label{eq:wordAdoptionReview}
   \end{equation}

Finally, to compute the percentage of adoption of a category $c$ associated with a set of reviews $R$, we computed the geometric mean of Eq.~\ref{eq:wordAdoptionReview}. Since our data may contain zeros, a constant value $k$ equal to the minimum adoption excluding zero has been added to each value in the set and later subtracted from the result.
   \begin{equation}
	\% adp(c) = ( \prod_{i=1}^{|R|}{ (\% adp(c, r_i) + k) } )^\frac{1}{n} - k
   \label{eq:wordAdoption}
   \end{equation}

In the above formula, $|R|$ is the cardinality of the set of reviews $R$. We always show results when $|R|>1$K reviews, so to have a percentage error less than 2\% with 95\% confidence interval~\cite{haukoos2005advanced,altman2013statistics}.

\paragraph{Metric operating on the dictionary lexicon.}

Beside the {\em adoption} metric defined in Eq.~\ref{eq:wordAdoption}, we define another metric called {\em term frequency gain}, which supports a more fine-grain level of investigation by operating at the lexicon level. Specifically, let $R_A$ and $R_B$ be two sets of reviews (e.g., reviews left in two given years and/or cities), let $r$ be a specific review belonging to $R_A \cup R_B$. Let $W$ be the set of words (unigrams) contained in $R_A \cup R_B$. For each word $w \in W$, we compute the logarithmically scaled term frequencies $tf_A(w)$ and $tf_B(w)$ associated with, respectively, $R_A$ and $R_B$. Finally, we compute the term frequency gain of each word $w$ as: 

\begin{equation}
tf_{gain}^{A/B}(w) = \frac{tf_A(w)}{tf_B(w)} 
\label{eq:tf_gain}
\end{equation}

Note that, because each term frequency in Eq.~\ref{eq:tf_gain} is normalised in [0,1], this metric allows us to detect words that are over-used in $R_A$ compared to $R_B$ ($tf_{gain}^{A/B}>1$), and vice versa ($tf_{gain}^{A/B}<1$).

\subsection{Validating our dictionary}
\label{subsec:validatingDict}

%\notelc{do we operate at sentence or review level here? needs clarification throughout 4.3}

To gain confidence in the ability of our dictionary and metrics to genuinely distinguish reviews that semantically belong to different categories, we performed two tests, one using a small set of manually labelled sentences, and one using the whole corpus of 3.2M unlabelled reviews.

\paragraph{Validation 1 -- Labeled reviews}

For the first validation test, we used the 1,868 manually labelled sentences at {\em step~3} above. For each sentence, we computed its business adoption and social adoption (category level-1) values, using Eq.~\ref{eq:wordAdoption}. We then compared these values to the  manual classification of such sentences performed by crowd-workers. Table~\ref{tab:BusinessVsSocialDict} shows the adoption of the business and social categories for the 1,868 manually annotated (ground truth) sentences. Let us consider the adoption of the business category first. As expected, the metric is much higher when computed over the business set than when computed over the social set (20\% against 3\% -- a decrease of -85\%). Conversely, the adoption of the social category is substantially higher when computed on the social set of reviews rather than the business set (at 10\% compared to 2\%, an increase of +400\%). This result is preliminary evidence that our dictionary and metrics are able to correctly distinguish the two level~1 categories. 

%\begin{table}[!ht]
%   \centering
%   \mytablesize 
%   \begin{tabular}{ ccc||c }   
%   \em               & \em Business set & \em Social set   & \em Relative change\\
%   \hline \hline
%   Business adoption &       20\%      &  3\%          &  -85\% \\
%   Social adoption   &        2\%      & 10\%          & +400\% \\
%   \hline \hline
%   \end{tabular}
%   \caption{Business and social adoption in our corpus of 1,868 manually annotated sentences having high agreement}
%   \label{tab:BusinessVsSocialDict}
%\end{table}

\begin{table}[!bht]
   \centering
   \mytablesize 
   \begin{tabular}{ ccc }   
   \em               & \em Business set & \em Social set \\
   \hline \hline
   Business adoption &       20\%      &  3\%            \\
   Social adoption   &        2\%      & 10\%            \\
   \hline \hline
   \end{tabular}
   \caption{Business and social adoption in our corpus of 1,868 manually annotated sentences}
   \label{tab:BusinessVsSocialDict}
\end{table}

Table~\ref{tab:BusinessVsSocialDict} also shows that the highest adoption of the business category is twice as high as the highest adoption of the social category (when computed over the business and social sets respectively). One may question whether this simply derives from the fact that the business vocabulary used by Airbnb guests (287 specific words concerning the property, location of the property, and professional conduct of the host) is substantially wider than the social one (68 specific terms concerning the social interaction between guest-host). To investigate whether this is indeed the case, we restricted the business and social lexicon to have the same number of words (i.e, we kept in our lexicon only the top $n$ words according to their term frequency for the business and social categories, with $n = \{10, 20, 40\}$). We found the exact same trend for all $n$. We take this as indication that guests' reviews are genuinely more prone to contain more business terms than social ones. 

%\notegq{I've removed the definition of relative change. I think it was confusing to introduce here a new metric. Also, when used in the paper, we always say ``change of adoption relative to the class ...'' which to me is intuitive and does not need a definition here.}

%This may derive from the fact that the business vocabulary used by Airbnb guests (287 specific words concerning the property, location of the property, and professional conduct of the host) is substantially wider than the social one (68 specific terms concerning the social interaction between guest-host). To take into account this difference of scales, in the rest of the paper we discuss results by quoting both the original adoption metric, and the the relative change of the adoption against a reference class. For example, in Table~\ref{tab:BusinessVsSocialDict}, the relative change of business adoption of the business set compared to the (reference class) social set is: $\frac{3\% - 20\%}{20\%} \times 100 = -85\%$.

%\notedq{ i m not sure i fully agree with this metric. in theory, the closest version should be $\%relative(c)=  \frac{\%adp(c) - \%adp(\bar{c})}{\%adp(c)} \times 100 =  \frac{20-3}{20} \times 100= + 85\%$. yet, i would still find it difficult to understand the meaning of this metric. simpler alternatives are  1) $(\%adp(c) - \%adp(\bar{c}))$ = indicates how far the actual value is from the null value; or 2) $(\frac{\%adp(c)}{\%adp(\bar{c})})$ = indicates the number of null value units  the actual value is made of}

\paragraph{Validation 2 -- Unlabeled reviews}

For the second validation test, we used unlabelled data and, specifically, all the 3.2 million guests' reviews. Airbnb  guests can choose to rent `whole apartments' as well as `shared/private rooms'. We expect to have more social interactions between host-guest when guests rent `shared/private rooms', compared to those occurring when guests rent `entire home/apt'. Therefore, we also expect  that reviews associated with `shared/private rooms' contain more social  terms than reviews associated with `entire home/apt'. To verify whether our dictionary and metrics can capture this intuition, we grouped reviews according to the type of property listed (i.e., `shared/private rooms' vs. `entire home/apt') and applied Eq.~\ref{eq:wordAdoption} to each set of reviews.

Table~\ref{tab:BusinessVsSocialDictByRoomType} shows the change of adoption of the  two level-1 categories in our dictionary, for  `shared/private rooms' relative to the reference class `entire home/apt'. We observe a slight decrease of adoption for the business category (from -11\% in Great Manchester, to -38\% in San Francisco) and a boost of adoption for the social category (from +76\% in San Francisco, to +209\% in Melbourne) when shifting from `entire home/apt' to `shared/private rooms'. This finding meets the intuition that reviews written for shared/private rooms discuss less business-related topics and more social-related topics than reviews written for entire apartments. We take this as further confirmation of the reliability of our dictionary.

\begin{table}[!ht]
   \centering
   \mytablesize 
   \begin{tabular}{ lccc }   
   \em Relative                 & \em Great      & \em        & \em           \\
   \em change of ...            & \em Manchester & \em London & \em Melbourne \\
   \hline \hline
   ... business adoption & ~--11\% & ~--28\% & ~--20\%  \\
   ... social adoption   & +173\% & +107\% & +209\%  \\
   \hline \hline
   \\
   \em Relative                 & \em New York  & \em San       & \em        \\
   \em change of...             & \em City      & \em Francisco & \em Sydney \\
   \hline \hline
   ... business adoption & ~--19\%  & ~--38\% & ~--20\%  \\
   ... social adoption   &  +106\% & +76\%  & +150\% \\
   \hline \hline
   \end{tabular}
   \caption{Change of the business and social adoption for the set of reviews associated with `shared/private rooms', relative to the reference class `entire home/apt'}
   \label{tab:BusinessVsSocialDictByRoomType}
\end{table}

We next proceed to illustrate four examples of market research questions that one can perform using our dictionary and metrics.

\section{The social-business dichotomy}
\label{sec:linguistic}

We conduct four different investigations that aim to shed light onto the big debate of whether Airbnb is a social interaction \emph{vs.} business transaction platform. First, we operate at the level of  {\em categories} defined in our dictionary to analyze at different granularities how Airbnb is evolving over a period of 10 years and for 6 different cities (Section~\ref{sec:years}). Second, we zoom in at the level of {\em lexicon} defined in our dictionary to detect micro-variations in trends, once again over time and space (Section~\ref{sec:words}). By segmenting reviews even further, we then investigate whether the business-social dichotomy varies for different groups of hosts (Section~\ref{sec:segments}) and for properties located in different neighbourhoods within the same city (Section~\ref{sec:neighborhoods}).

\subsection{The dichotomy over the years}
\label{sec:years}

\begin{figure}[!ht]
    \centering
    \includegraphics[width=.49\textwidth]{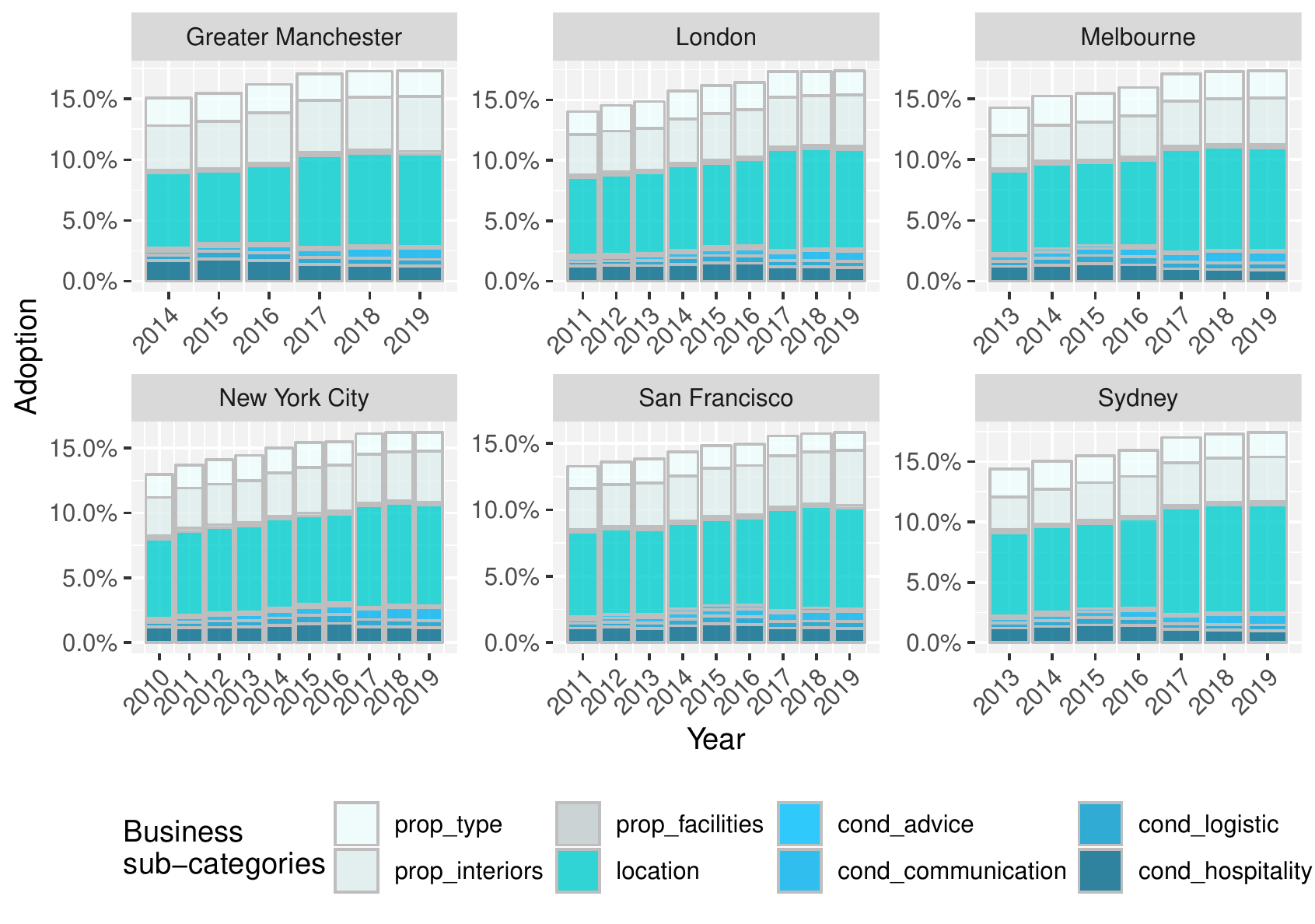}
    \includegraphics[width=.49\textwidth]{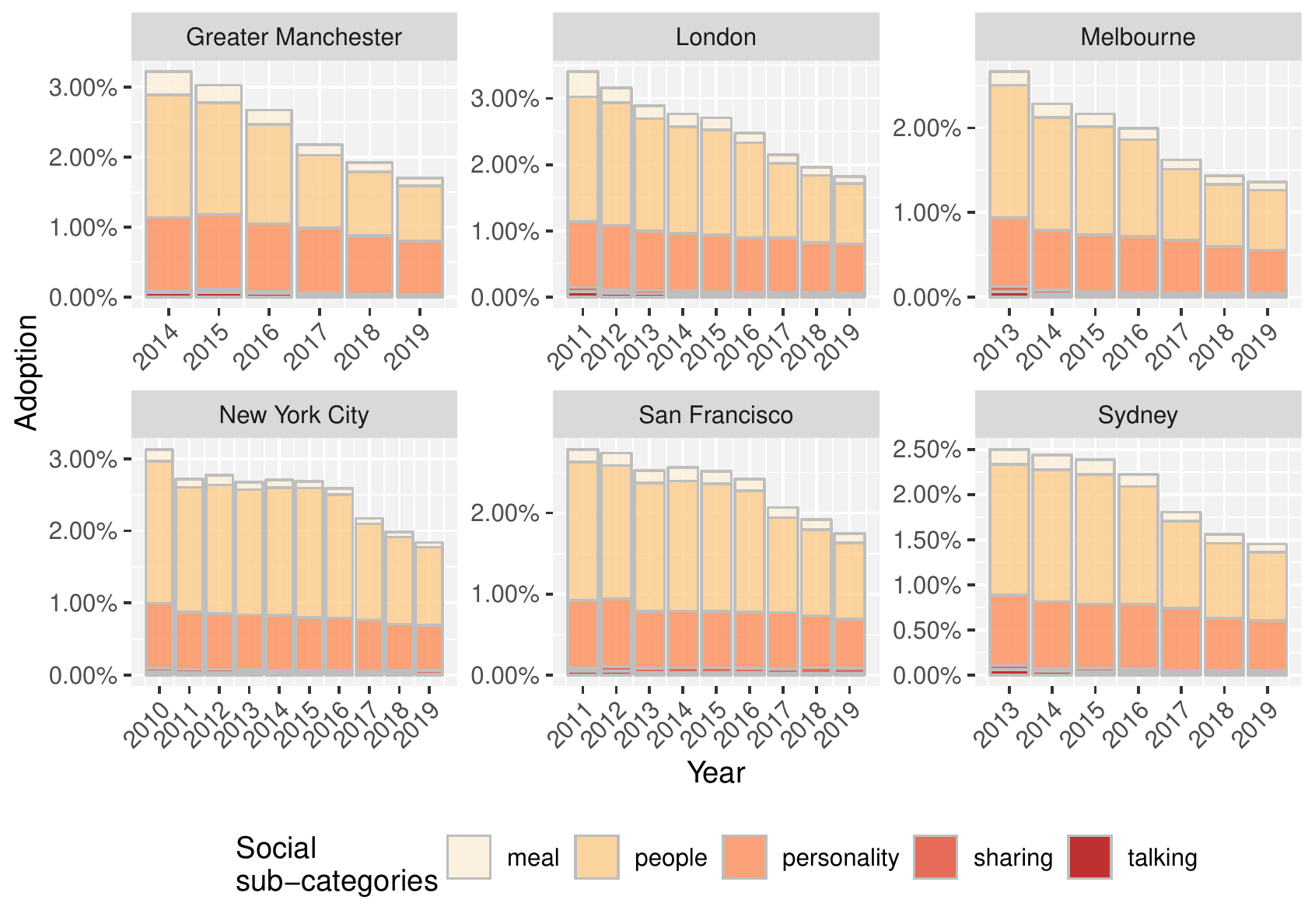}
    \caption{Adoption metric by year and city}
    \label{fig:ThemesVsYearByCity}
\end{figure}

To begin with, we investigate whether Airbnb is evolving as a platform where guests are more concerned with business aspects of the service or with social ones. We perform this analysis by grouping reviews on a per year and per city basis, and by computing the adoption metric defined in Eq.~2. Figure~\ref{fig:ThemesVsYearByCity} plots the adoption of the `business' and `social' level~1 categories across each year and for each analysed city; the different color shades of the plot show the adoption of the level~3 categories.

%Figure~\ref{fig:ThemesVsYearByCity} plots the results. \notelc{something needs be said about how to read the figure, once finalised. top is biz, bottom is social, shades are for level 2 etc. caption needs be fixed too}.

%\notedq{the reviewer might wonder why the levels for business and social are not the same}
%\notegq{I agree with Daniele. This weekend I'll make a new plot and I'll uniform levels. So, Fig.1 will report only level~3 categories for both business and social level~1 categories.}

Overall, we find that the adoption of the `business' category is increasing over time, while the adoption of the social category is steadily decreasing. For example, in London, the adoption of the business category in 2011 is 14\%, whereas it is 17.5\% in 2019 -- the  relative increase is of 25\% in a 9 year temporal window, with a growth of 2.8\% per year. Conversely, the adoption of the `social' category is 3.5\% overall in 2011, whereas it decreases to 1.9\% in 2019. This represents a reduction of 45\% in a 9 year temporal window, with a relative decrease of 5\% per year.  

To gain more fine-grained insights, we inspect Figure~\ref{fig:ThemesVsYearByCity} further, and observe trends within the `business' category. We find that, at level~3 categories, `location', `property type' and `interiors' are the most frequently discussed ones; as an example, in London they collectively gather 14.5\% of adoption in 2019, with a constant growth since 2011.  `Hospitality' is the level~3 category with highest adoption within the `business conduct' level~2 (sub)category; as an example, in London it reached its highest adoption level (around 1.5\%) in 2015--2016 with a consistent increase from 2011, but either stalled or even slightly decreased afterwards. If we now move our attention to the `social' category, we find that `people' and `personality'  are the most frequently discussed level~3 (subsub)categories, with about 1\% and 0.5\% adoptions in 2019, respectively. However, both of them exhibit a negative slope of adoption rate across all years. The other level~3 social categories, namely `meal', `sharing' and `talking', are rarely used in the whole observation period.  
%
%`property' and `location'  are the most frequent business (sub)categories and they collect 5\% of adoptions in 2019, with a constant growth since 2011. As for `business conduct', it reached its highest adoption level (around 2.5\%) in 2016 with a consistent increase from 2011, but either stalled or even slightly decreased afterwards. We can zoom in even further; for example, if we now move attention to the social category and its level~3 (subsub)categories, we find that `people' and `personality'  are the most frequently used in reviews, with about 1\% and 0.5\% adoptions in 2019, respectively. However, both of them exhibit a negative slope of adoption rates across all years. The other level~3 social categories, namely `meal', `sharing' and `talking', are rarely used in the whole observation period. 

Figure~\ref{fig:ThemesVsYearByCity} also shows that all the identified trends are confirmed for each city, despite the fact they are located in different countries/continents. This suggests that Airbnb language evolution is happening at a global scale, at least in Western countries.

\begin{figure}[!t]
    \centering
    \includegraphics[width=.3\textwidth]{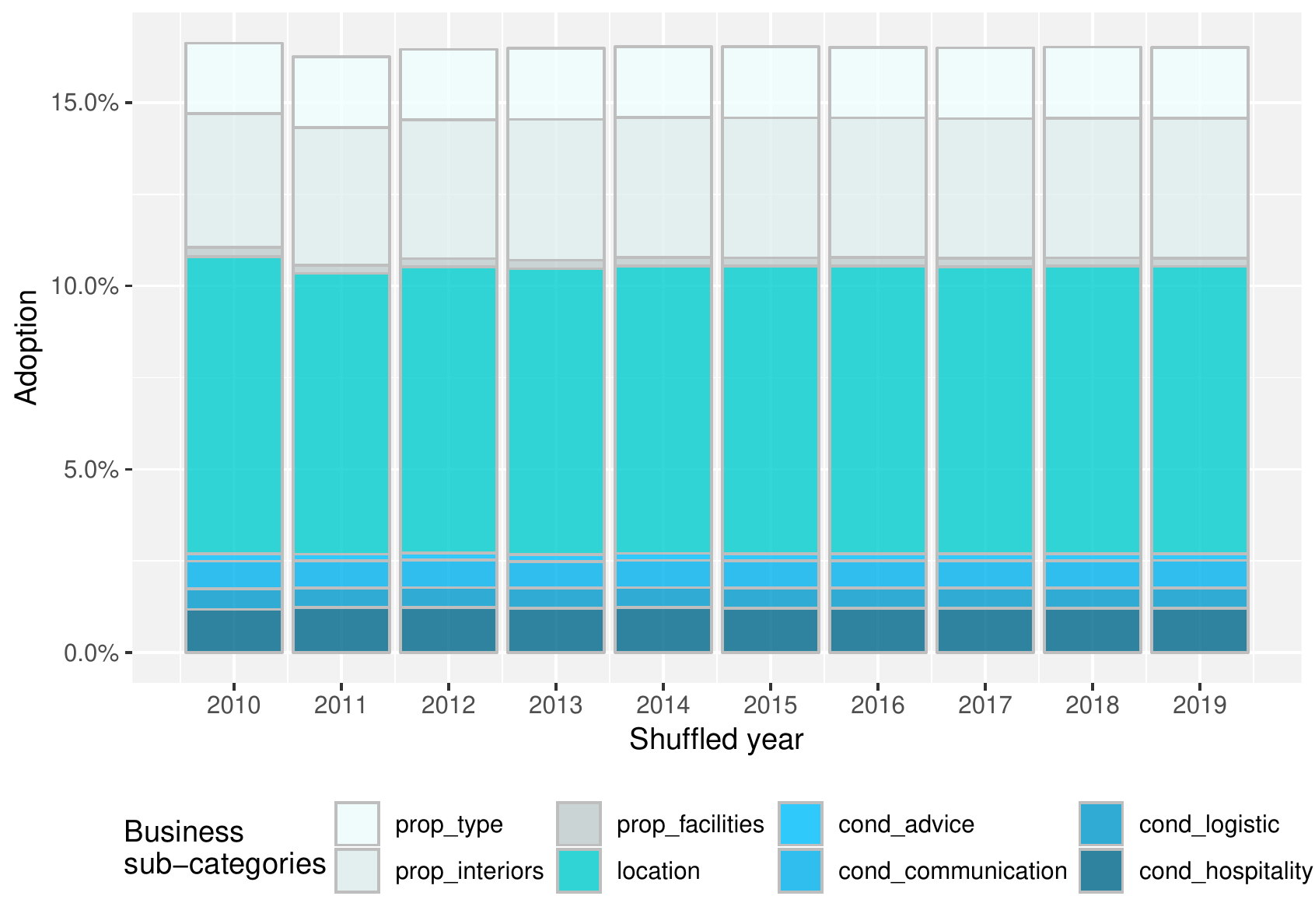}
    %\hspace{25mm}
    \includegraphics[width=.3\textwidth]{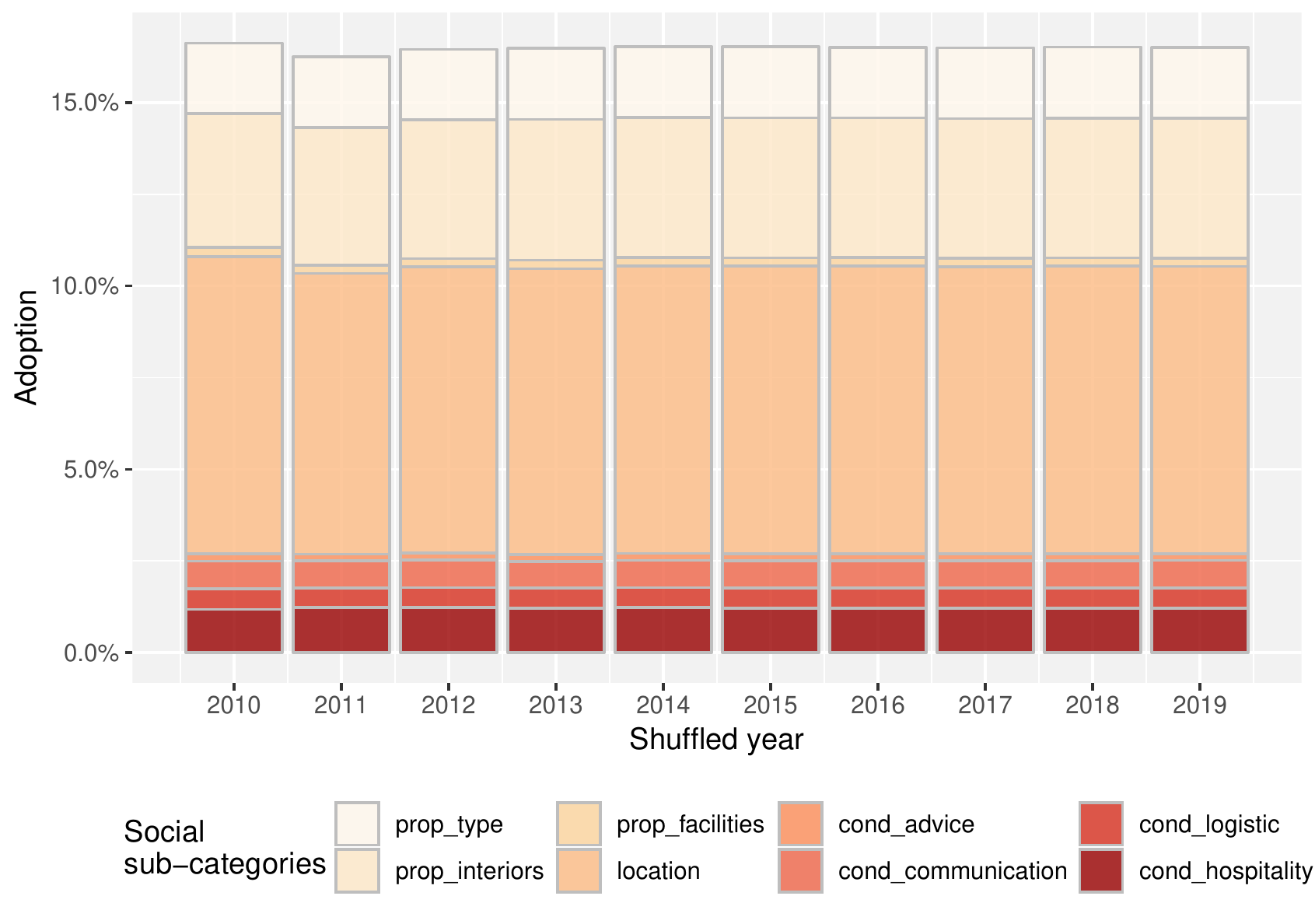}
    \caption{Adoption metric for the null model}
    \label{fig:NullSanityCheck}
\end{figure}

\paragraph{Sanity checks.} To gain confidence in the results presented above, we performed both a statistical validation of the proposed metrics and an analysis of potential  confounding factors. In terms of statistical validation, we built a null (random) model by shuffling the years in our dataset and then repeated the whole analysis on it. 
Figure~\ref{fig:NullSanityCheck} shows the adoption metric applied to the (random) null model averaged across all cities, since trends were found to be similar. We observe flat trends throughout.
%Figure~\ref{fig:NullSanityCheck} shows that the adoption metric (Equation~\ref{eq:wordAdoption}) exhibits flat trends in each city and for each category level. 
Furthermore, we compared the distribution of adoption of both business and social categories in the years from 2010 to 2012 against their distribution in the years from 2017 to 2019, by running the {\em Wilcoxon rank sum test}. The obtained {\em p-value} $< 2.2 \ e^{-16}$ confirms the difference in the mean value showed above is statistically significant. 

\begin{figure}[!thb]
    \centering
    \includegraphics[width=.49\textwidth]{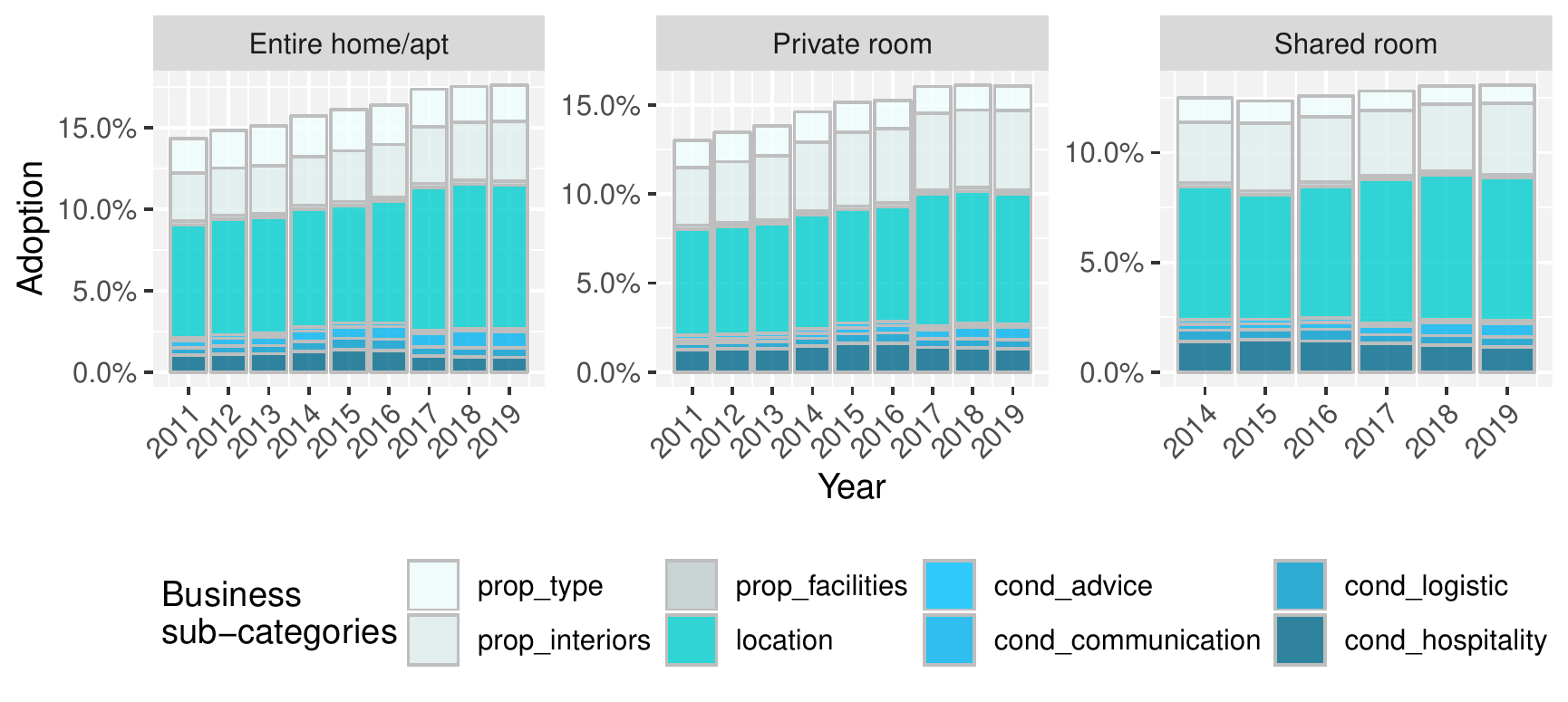}
    \includegraphics[width=.49\textwidth]{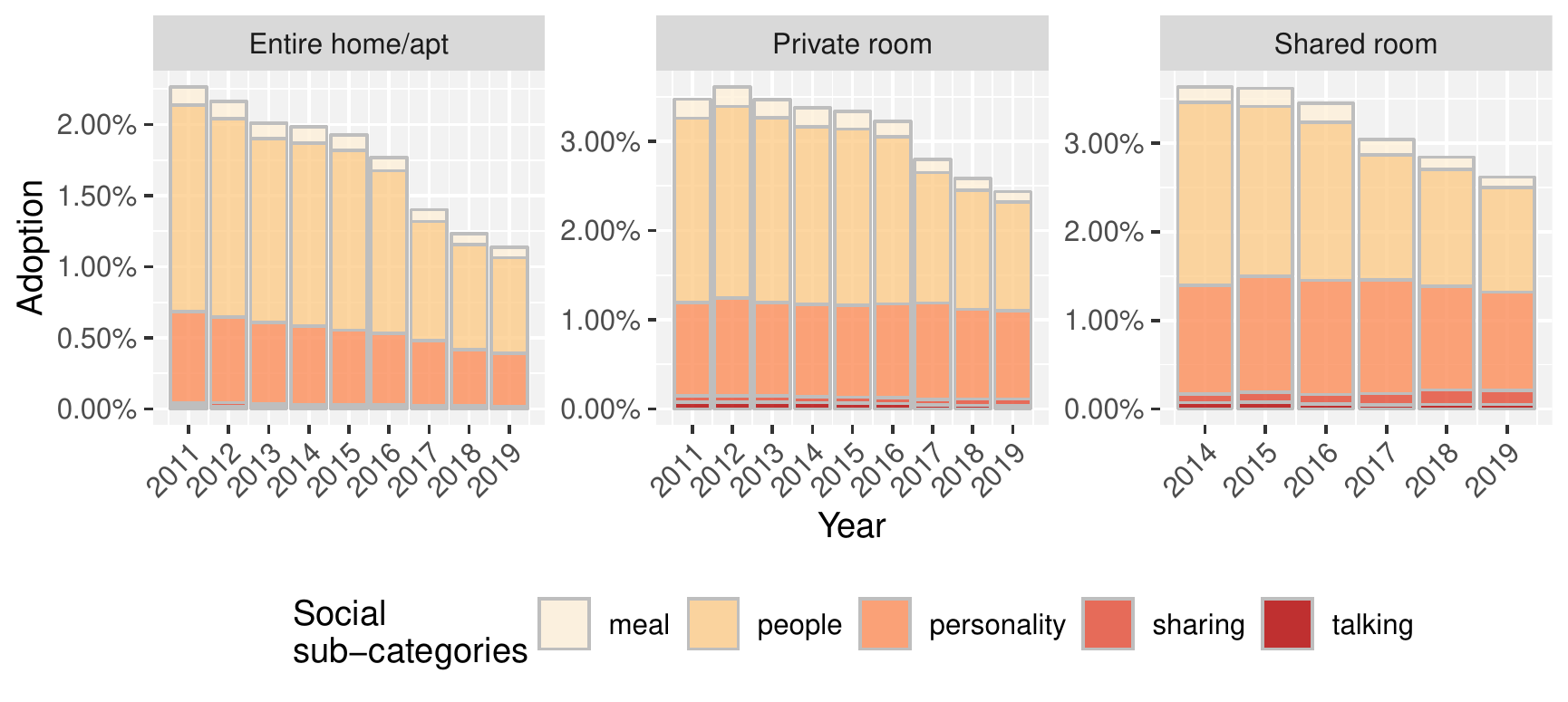}
    \caption{Adoption metric by year and room type}
    \label{fig:ConfoundingSanityCheck}
\end{figure}

To control for potential confounding factors, we considered both {\em review length} and {\em room type}. Intuitively, both of them could be a cause of the observed phenomenon: a recorded prevalence of short reviews in the late years could cause the reduction of the adoption of the `social' category; this is true also if the number of reviews associated with `entire apartments' as opposed to `shared/private room' increases drastically in the last years considered in our investigation. To exclude these confounding factors, we proceeded by binning the reviews in our dataset according to their length and room type. We re-plotted the `business'/`social' adoption for each year from 2010 to 2019 and for each bin. As an example, Figure~\ref{fig:ConfoundingSanityCheck} illustrates the adoption metric for the `business' and `social' categories across each year and across each value of room type (`entire home/apt', `private room', `shared room'). Results have been averaged by city since trends were found to be similar. The illustrated trends are consistent with those already shown in Figure~\ref{fig:ThemesVsYearByCity}, suggesting that the findings reported in this section are not a consequence of these confounding factors.
%Figure~\ref{fig:ConfoundingSanityCheck} illustrates trends that are consistent with those already shown in Figure~\ref{fig:ThemesVsYearByCity}, suggesting that the findings reported in this section are not a consequence of these confounding factors. \notelc{the above needs be checked once final charts are in}

\subsection{The dichotomy in words}
\label{sec:words}

The above analysis suggests that Airbnb is evolving as a platform where guests are more concerned with business aspects, rather than with social aspects, of the hospitality. We can examine more precisely what aspects of the service are behind this trend by operating at the level of the lexicon and computing the term frequency gain metric (Eq.~3). As an example, we binned reviews so to cover a late period (set $A$, reviews written between 2017--2019) and an early one (set $B$, reviews written between 2010--2012). For each review belonging to the periods above, we consider each word (unigram) $w$ and compute the {\em term frequency gain} $tf_{gain}^{A/B}$ as defined in Eq.~\ref{eq:tf_gain}. This metric allows us to detect words that are over-used in the late period compared to the early one ($tf_{gain}^{A/B}>1$), and vice versa ($tf_{gain}^{A/B}<1$). To ease the analysis, we discarded very unpopular words, i.e., words having a total term frequency lower then $10^{-5}$. 

\begin{figure}[!th]
    \centering
    \includegraphics[width=.5\textwidth]{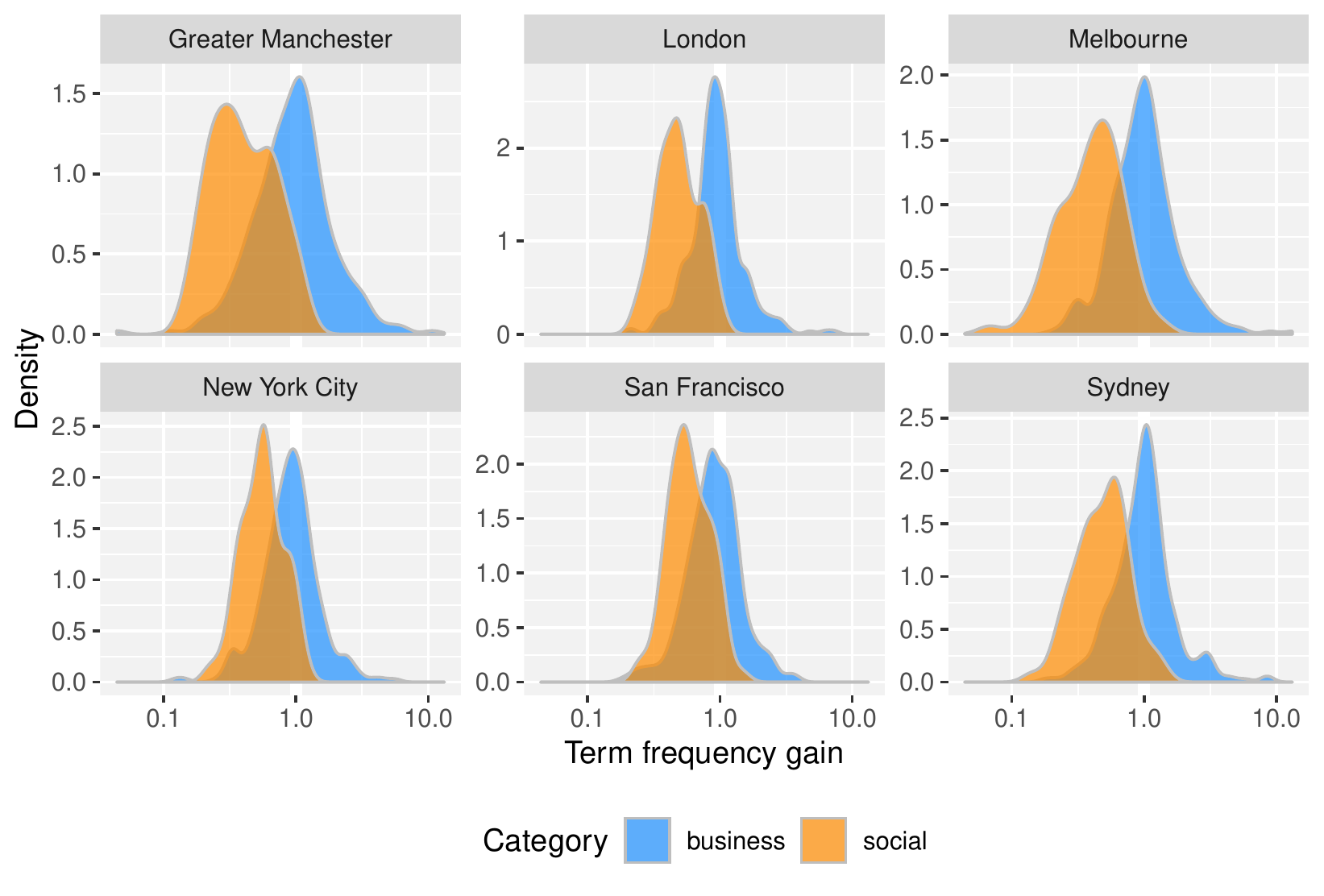}
    \caption{Density distribution of $tf_{gain}^{A/B}$ of each category in each city}
    \label{fig:CountThemeVsGainByCity}
\end{figure}

We start this investigation by plotting the density distribution of this metric for the two broadest categories of our dictionary (`business' and `social'). Figure~\ref{fig:CountThemeVsGainByCity} shows the results for each analysed city and highlights two interesting trends. First, the `business' category has words associated with both positive and negative term frequency gain. This means that some aspects of the `business' category are indeed over-emphasised in the late period compared to the early years when Airbnb was a young service; however, there are also `business' words which experience an opposite trend. Second, the great majority of words that are part of our `social' lexicon are under-used in the late period compared to the early years when Airbnb was a young service. This is valid in all cities under study.

\begin{figure}[!ht]
    \centering
    \includegraphics[width=.5\textwidth]{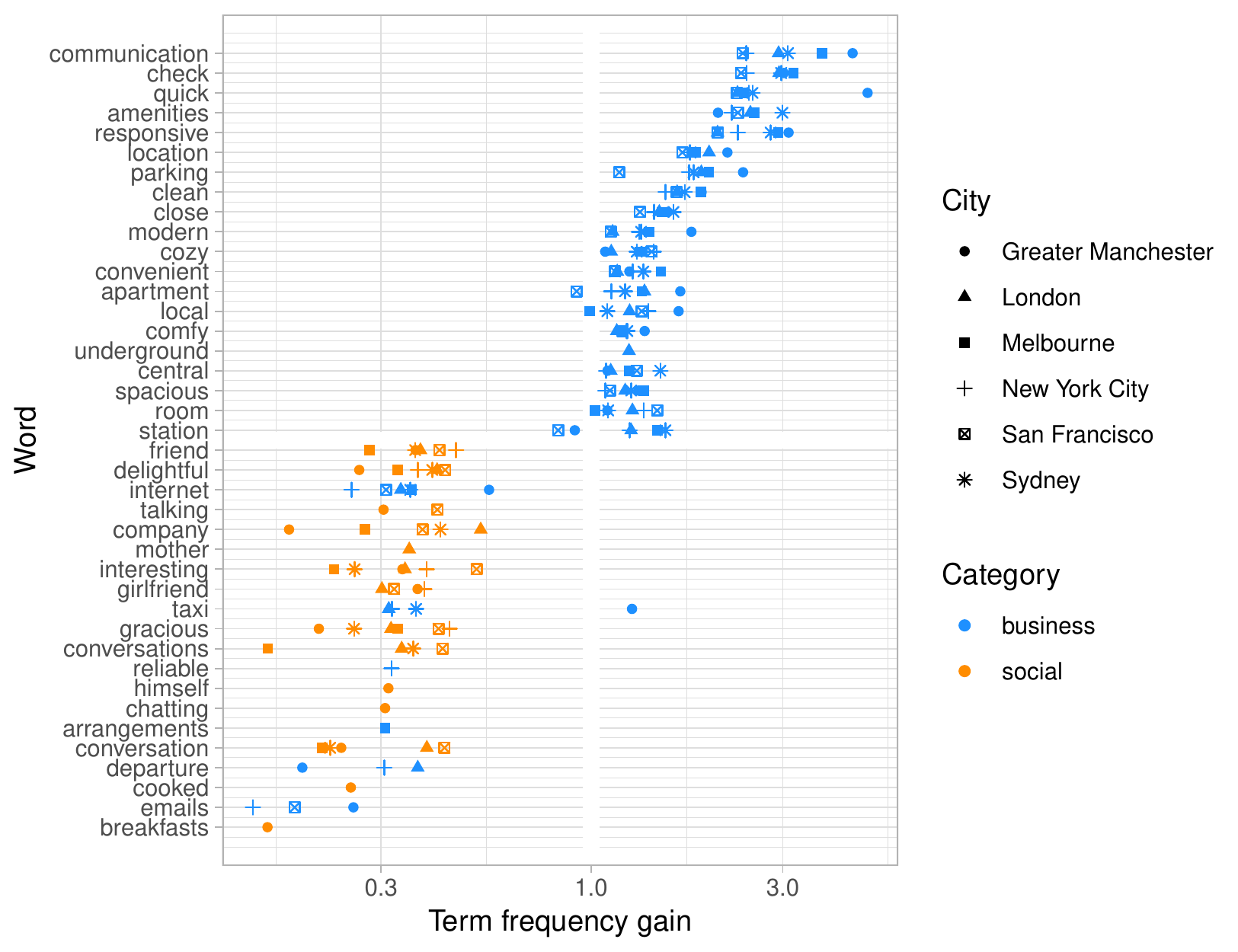}
    \caption{Words contained in our dictionary having the top/bottom 20 term frequency gain}
    \label{fig:GainVsWordByCity}
\end{figure}

Figure~\ref{fig:GainVsWordByCity} shows the top-20 words in our dictionary that exhibit the strongest decline/increase of term frequency gain (computed as average across all cities). As one would expect based on our findings so far, all the top-20 words belong to the `business' category; interestingly, we observe that it is the `location' (sub)category that mostly drives this trend, with words such as `parking', `local', and `central'. Conversely, 70\% of the bottom-20 words belong to the `social' category, and these words span different social (subsub)categories, from `personality' (e.g., `gracious'), to `people' (e.g., `friend'), to talking (e.g., `delightful', `company', `conversations'). 

\subsection{The dichotomy across market segments}
\label{sec:segments}

\begin{figure}[!th]
	\centering
	\includegraphics[width=.5\textwidth]{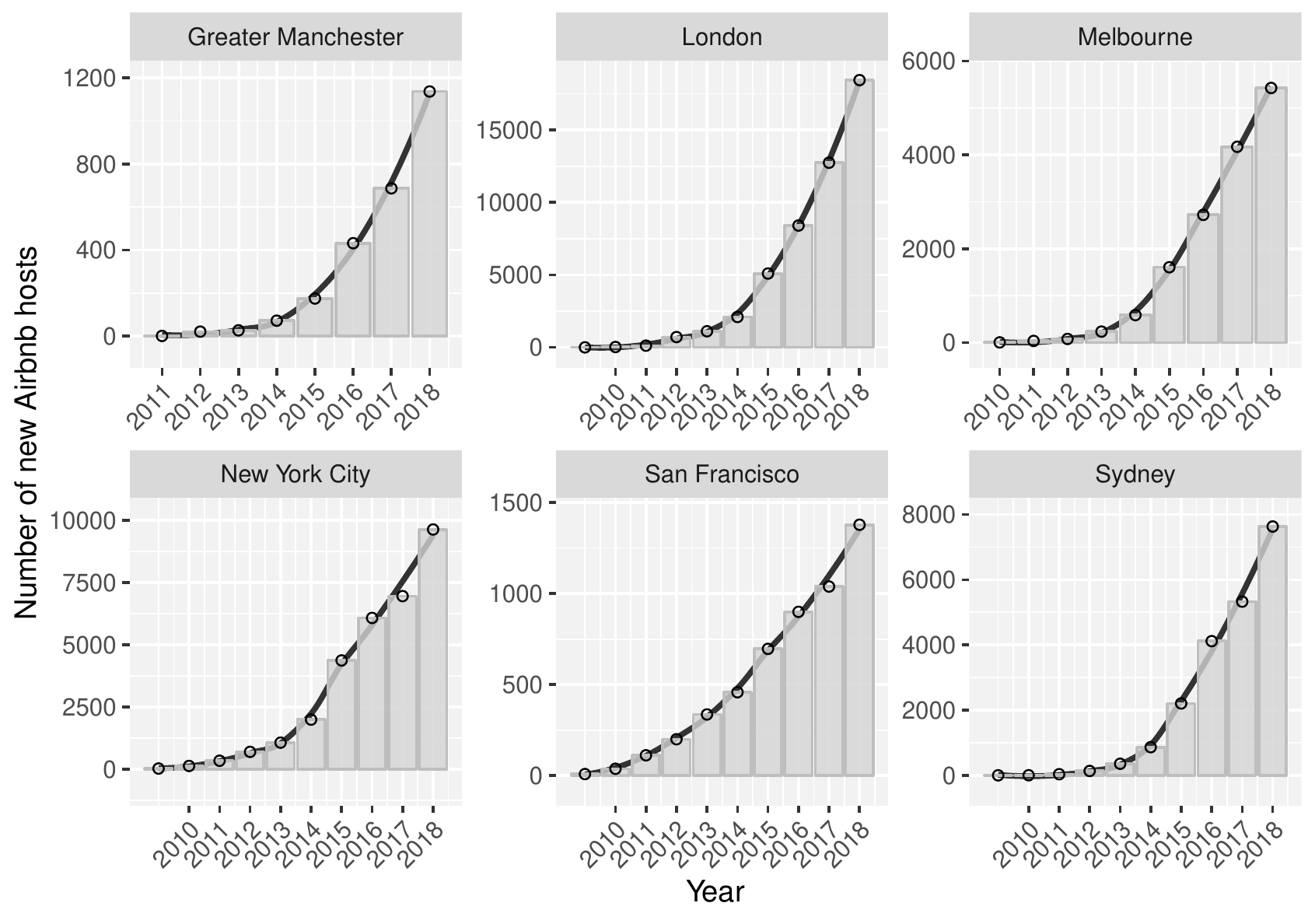}
	\caption{Adoption of Airbnb in each city} 
	%\notedq{is the y-label right? it might be number of airbnb users (not only the new ones)} }
	%\notelc{importante, da verificare}
	%\notegq{Ho verificato, il grafico della technology adoption riporta quando un utente fa join della piattaforma, per cui e' un nuovo utente. A meno che non mi sfugge qls, penso sia corretto il grafico.
	\label{fig:adoptionbByYear}
\end{figure}

By segmenting reviews not only by city and year but also by host (i.e., service provider) characteristics, our approach enables investigations across different market segments. As an example, in this section we  segment hosts  based on the concept of technology adoption~\cite{rogers2010diffusion,hage1980theories,van2000research,zaltman1973innovations}. In the literature, users are classified as: {\em innovators} (first 2.5\% of users) adopting a new technology, {\em early adopters} (subsequent 13.5\% of users), {\em early majority} (the following 34\% of users), {\em late majority } (34\% of remaining users), and {\em laggards }(last 16\% of users). To understand the current Airbnb adoption era, we computed the number of new users for each year and for each city involved in our analysis; we plot the corresponding results in Figure~\ref{fig:adoptionbByYear}.

By comparing the shape of the obtained curves with the Gaussian-shaped one reported in the related literature~\cite{rogers2010diffusion,van2000research,zaltman1973innovations}, we  conclude that Airbnb has not reached the late majority phase yet, and this result is homogeneous across all the cities in our investigation. Moreover, by observing the adoption rates, we hypothesise that Airbnb is in the middle of the early majority phase. Following this reasoning, we  segment hosts in our dataset in: innovators (first 5\%), early adopters (subsequent 45\%), and early majority (remaining 50\%). We then linguistically analyse the reviews that each such bin collected. 

For ease of presentation, we focus this analysis on the two level-1 categories only, and present results in a concise way by means of what we call `social score':
   \begin{equation}
	social score = z(\% adp(social)) - z(\% adp(business)),
   \label{eq:zscore}
   \end{equation}
where $z(\% adp(social))$ and $z(\% adp(business)$ are, respectively, the z-scores of the `social' and the `business' adoption on a set of reviews. Note that, since both $z(\% adp(social))$ and $z(\% adp(business))$ are normalised and unitless numbers, Eq.~\ref{eq:zscore} is able to directly compare the `social' and `business' adoption despite their difference of scale, thus telling us whether a particular set of reviews is biased towards the former or the latter category.

%We find that, in all analysed cities, innovators have associated reviews containing less business terms \notelc{where is this shown? why do we talk abut terms and not social score? what metrics are we using in this analysis? confusing} and more social ones than the other categories of users. As an example, in Great Manchester, innovators have associated reviews having 15\% of business terms, and 3.1\% of social terms. In the same city, the early majority of users have associated reviews having 17.5\% of business terms (17\% more than innovators), and 1.75\% of social ones (43\% less than innovators). Once again, this result is confirmed for each city in our investigation, as shown in Figure~\ref{fig:zscoreVsTechByYear} \notelc{the figure does not differentiate by city? clarify} reporting the social score defined in Equation~\ref{eq:zscore} for the reviews received by the three categories of hosts identified above and for each year. Figure~\ref{fig:zscoreVsTechByYear} shows that in each year innovators receive reviews having an higher social score than the other categories of users do. 

\begin{figure}[!t]
	\centering
	\includegraphics[width=.5\textwidth]{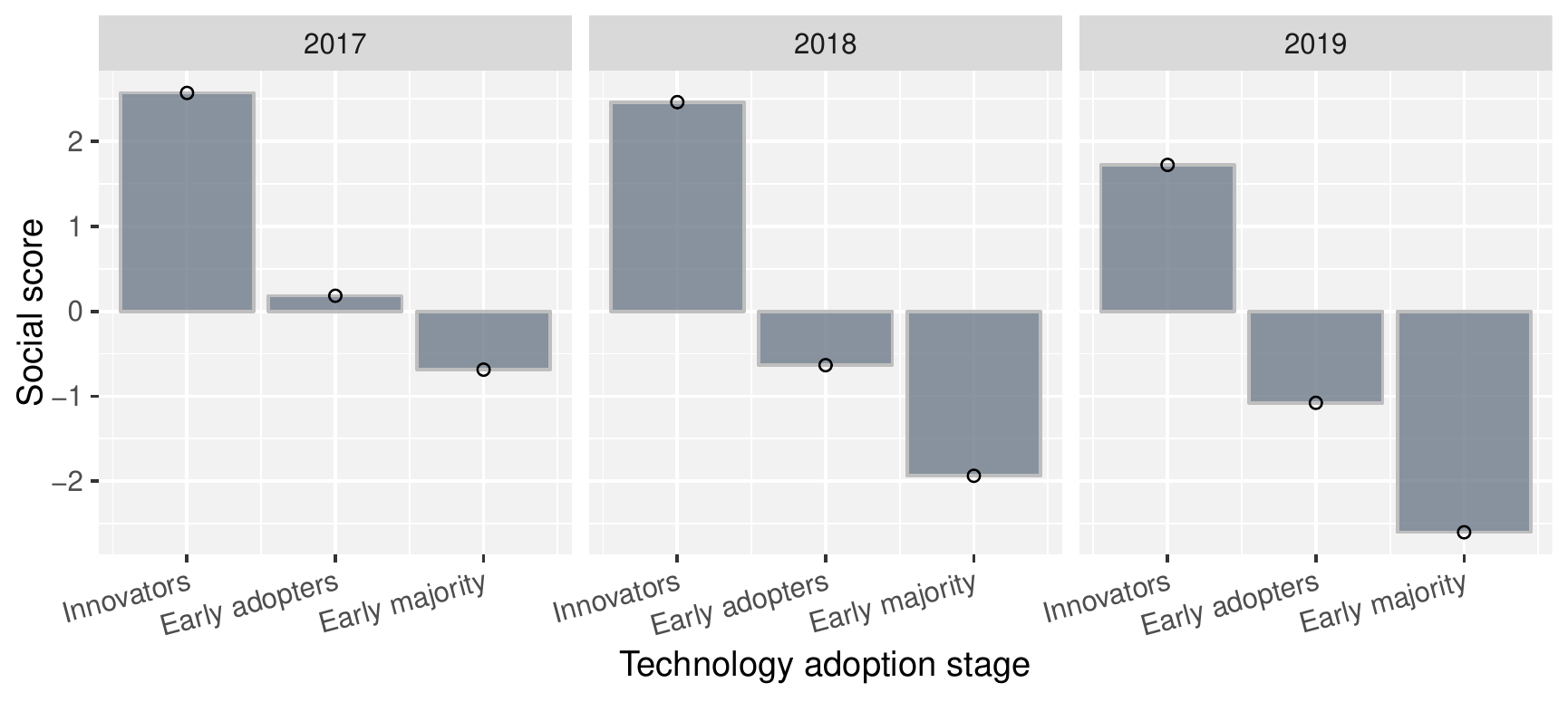}
	\caption{Social score against Airbnb technology adoption}
	\label{fig:zscoreVsTechByYear}
\end{figure}

Figure~\ref{fig:zscoreVsTechByYear} shows results averaged across all cities, since trends were found to be similar. We observe that, in each year, innovators receive reviews with a higher social score than the other categories of hosts do. We speculate that innovators may engage more in social interactions (as the original sharing economy manifesto wanted), and thus receive reviews with higher social scores. Figure~\ref{fig:zscoreVsTechByYear} also indicates that the percentage of social reviews received by innovators is decreasing over time. This result is coherent with a platform adaptation phenomenon, and may suggest that, even though innovators remain overall more social than the other categories of users, they are undergoing an adaptation process following the evolution of Airbnb towards a more business-oriented model. After controlling for the same confounding factors discussed in the previous section, we find that neither review length nor room type affected our results.

\subsection{The dichotomy across neighbourhoods}
\label{sec:neighborhoods}

As a final example of market research investigation one may perform, we illustrate how to segment reviews at a finer level of spatial granularity, so to investigate the possible presence of varying platform adoption dynamics within a single city. To illustrate how, we subdivided each city in its electoral districts, which are geographic areas of different size designed to have a similar number of residents. For each such area, we computed two scores: a social score, computed using Eq.~\ref{eq:zscore} over the set of reviews left for Airbnb properties located in such area; and an {\em Airbnb penetration rate}, computed as the ratio of the number of active Airbnb listings (that is, the number of listings receiving at least one review), over of the maximum number of listings in any given district, so to normalise such a rate between $[0,1]$. The latter has previously been found to be a good proxy for central / tourist areas~\cite{quattrone2016airbnb}.
%; here we will use this rate to detect areas with high guest turn-over, thus potentially helping identify areas more at risk of becoming ghost neighbourhoods.

%\notelc{ho rimosso sta frase che non fa fit, e' troppo specularive: Recent studies have suggested that areas with very high Airbnb penetration are rapidly becoming ghost neighbourhoods, eroding local communities and compromising people security.\footnote{https://www.theguardian.com/commentisfree/2018/feb/12/profiteers-killing-airbnb-erode-communities} The ability to perform quantitative linguistic analysis of Airbnb reviews at the level of neighbourhood, as opposed to the whole city, would afford us the ability to contribute data-driven insights to such studies. }

\begin{figure}[!t]
	\centering
	\includegraphics[width=.5\textwidth]{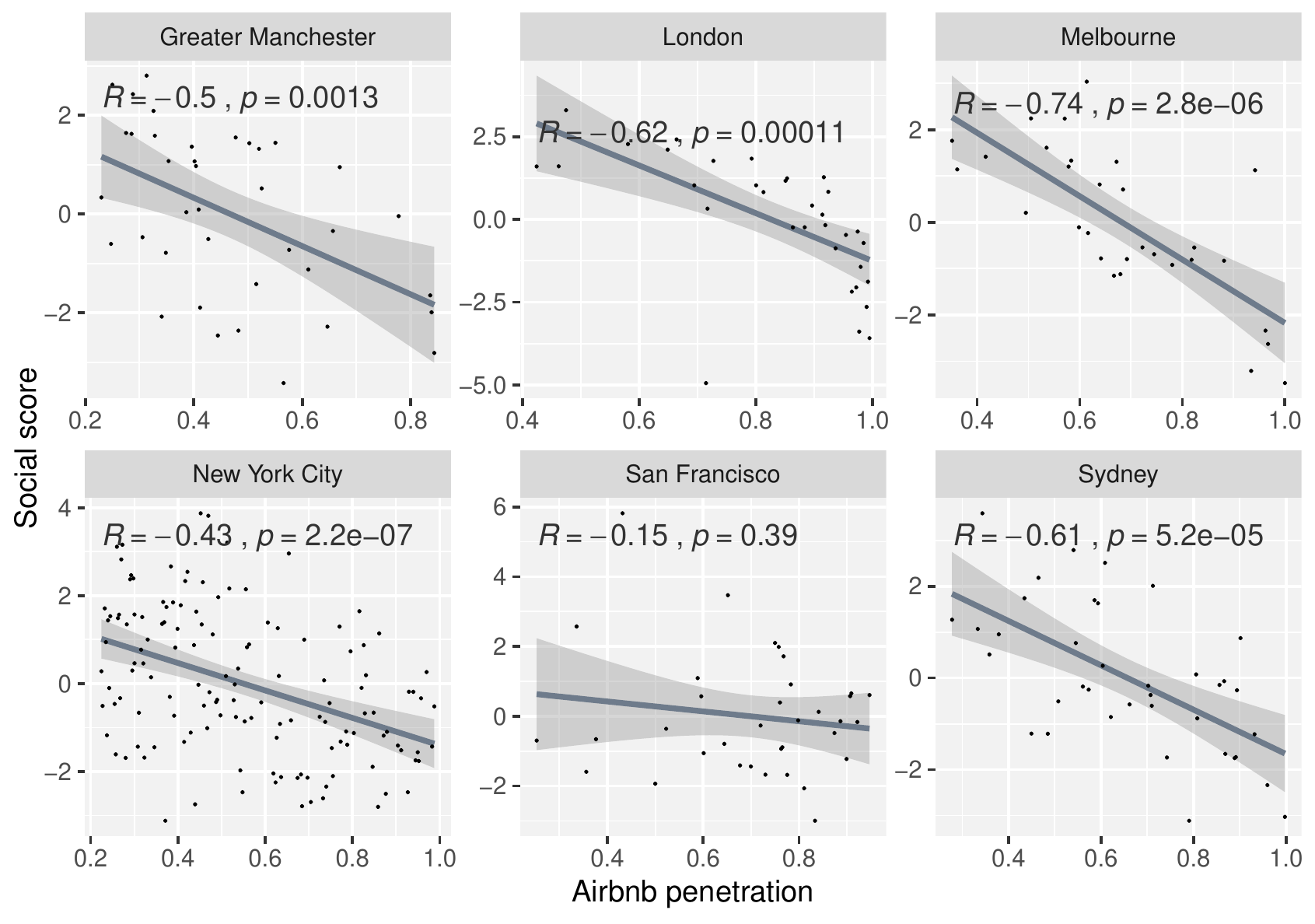}
	\caption{Social score against area Airbnb penetration rate (on a per city basis)}
	\label{fig:ScoreVsAirbnbByCity}
\end{figure}

Figure~\ref{fig:ScoreVsAirbnbByCity} shows the scatter plot (along with Pearson Correlation) between the Airbnb penetration rate and the social score for neighbourhoods in each city in our dataset. We observe that neighbourhoods with very high Airbnb adoption rates show lower social scores than those with lower penetration rates (Pearson correlation up to -0.74). Results are valid across all  cities considered, excluding San Francisco where the correlation is not statistically significant ($p$-value $>0.01$). These results would suggest that the Airbnb hospitality service is being valued more from a business point of view in central and tourist areas, whereas the social element is to be found once we move towards off-the-beaten track areas. 

%\notelc{so what? needs elaboration.} \notegq{I agree. I don't know how to elaborate though. I have two ideas. (1) We may say that orthogonal studies can be performed by crossing these results with official census data. This would allow us to infer the socio-economic demographics of social and business hosts. This would enrich the understanding how different demographic groups behave in the Airbnb platform. (2) We may use this information to build a recommender system. Users may be recommended to areas according to their needs (their business or social interests). Does it make sense?}\notelc{too speculative, removed and left for discussion in next section}

\section{Discussion and Conclusion}\label{sec:conclusion}

According to the `crossing the chasm' theory~\cite{moore1999crossing}, there are five segments of technology adoption: the first three segments `innovators', `early adopters', and `early majority'  (which together make the chasm) are followed by `late majority' and `laggards'. Based on our results, Airbnb has crossed the chasm by moving its adoption from the innovators  and early adopters to the larger market segment of the early majority, and is still expected to grow in this segment.

With the arrival of the early majority, norms have changed though: early adopters engaged more in social interactions than the early majority are currently doing. Given this majority, Airbnb now revolves around business-focused experiences rather than socialisation. That is surprisingly consistent across the various countries our six Western cities are located in.  Interestingly, the behaviour of early adopting hosts does not apply to early adopting areas: despite tourist areas being the initial ones to be offered on the platform~\cite{quattrone2016airbnb}, they are predominantly engaged in business-oriented experiences compared to properties offered in less central areas. 

These findings were made possible thanks to a platform (Airbnb) specific dictionary that we constructed, starting from ready available guests' reviews and using a combination of thematic analysis and machine learning techniques. Using this dictionary and a few metrics we defined, it is now possible to perform quantitative linguistic analysis of Airbnb experiences at scale, and use the findings to inform strategic business decisions that span different directions. For example, {\em (i) improved guest experience}: Airbnb  developers could add functionalities to the platform, such as guest/guest recommender systems, to connect like-minded people based on the topics they discuss in their reviews. Furthermore, rather than going for a `one size fits all' business model, they may try to leverage the platform diversification (in peer composition and district offerings), and enrich the service  offered with new features tailored to a guest's willingness to socialise by, for example, preferentially ranking hosts among the early adopters or properties in less tourist areas. {\em (ii) Improved host experience}: Airbnb could offer new hosts information and online training on how to attract guests, for example by recommending they offer services/experiences that guests to that city care the most about; as changes in what guests care about are detected, the platform can offer up-to-date vetting and training to its hosts, so to maintain guests' satisfaction high. {\em (iii) Tailored marketing}: by knowing what guests value, Airbnb can create hyper local advertising campaigns, for example highlighting more the efficiency of the service rather than its hospitality, to appeal to certain market segments that can vary by geographic location and over time. {\em (iv) Data-driven regulation}: by knowing the local market position of Airbnb, authorities and platform owners can co-create policies that differentiate business/leisure travels. 

Our dictionary and metrics can also support social science researchers in their investigations. For example, despite analysing cities in different countries and continents, we acknowledge that this work has so far been restricted to the Western world. We thus cannot answer questions of globalisation and platform adaptation that expand beyond it. Future studies should be conducted both in Eastern countries and in developing ones, both of which have largely been neglected by the current sharing economy literature, partly because of a lack of scalable analytical tools~\cite{Dillahunt:2017}. As we do so, our dictionary may have to me amended, or new ones may have to be developed, so to properly analyse Airbnb reviews in a new geographic context and/or in a different language. One can automatically assess whether our dictionary is valid (e.g., in a new city, at a future point in time, in a different - translated - language) by quantifying the proportion of words modelled by our dictionary, with the expectation that the same dictionary can be used for a number of years within cities sharing similar cultural traits. When a new custom dictionary needs to be built, the very same inductive approach proposed in this paper can be reproduced. This process offers great scalability advantages compared to the traditional market analysis approaches based on interviews, with the expectation that the effort to employ (thousands of) crowd-workers to conduct a new thematic analysis to be significantly less than that to recruit and interview (tens of) Airbnb users (with domain experts needed in both cases).
 
In addition to replicating our analysis, new lines of investigation can be pursued, delving deeper into peer segmentation, for example to explore questions of gender-specific and age-specific values in the Airbnb hospitality service. Last but not least, using a similar approach to the one proposed in this paper, one could go beyond the business-social dichotomy and develop dictionaries that enable orthogonal explorations, for example on the theme of trust, not least because trust is one of the main currency in  the sharing economy. Nowadays, given Airbnb's focus on business at the price of socialisation, cultivating trust might not be a priority and, as such, trust deficits might stand in the way of growth -- the very same growth that could help Airbnb decisively move well beyond the chasm.

\bibliographystyle{plain}
\bibliography{www2020}

\begin{thebibliography}{10}

\bibitem{alsudais2019large}
Abdulkareem Alsudais and Timm Teubner.
\newblock Large-scale sentiment analysis on airbnb reviews from 15 cities.
\newblock 2019.

\bibitem{altman2013statistics}
Douglas Altman, David Machin, Trevor Bryant, and Martin Gardner.
\newblock {\em Statistics with confidence: confidence intervals and statistical
  guidelines}.
\newblock John Wiley \& Sons, 2013.

\bibitem{bancoro2018tourists}
Amita Inah~Marie Bancoro and Adarsh Batra.
\newblock Tourists’ motivation in using sharing accommodation in bangkok.
\newblock In {\em Global Conference on Business, Hospitality, and Tourism
  Research (GLOSEARCH 2018)}, 2018.

\bibitem{boros2018airbnb}
Lajos Boros, G{\'a}bor Dud{\'a}s, Tam{\'a}s Kovalcsik, S{\'a}ndor Papp,
  Gy{\"o}rgy Vida, et~al.
\newblock Airbnb in budapest: analysing spatial patterns and room rates of
  hotels and peer-to-peer accommodations.
\newblock {\em GeoJournal of Tourism and Geosites}, 21(1):26--38, 2018.

\bibitem{braun2006using}
Virginia Braun and Victoria Clarke.
\newblock Using thematic analysis in psychology.
\newblock {\em Qualitative research in psychology}, 3(2):77--101, 2006.

\bibitem{bridges2018if}
Judith Bridges and Camilla V{\'a}squez.
\newblock If nearly all airbnb reviews are positive, does that make them
  meaningless?
\newblock {\em Current Issues in Tourism}, 21(18):2057--2075, 2018.

\bibitem{cheng2019airbnb}
Mingming Cheng and Xin Jin.
\newblock What do airbnb users care about? an analysis of online review
  comments.
\newblock {\em International Journal of Hospitality Management}, 76:58--70,
  2019.

\bibitem{cuong2019eliminating}
Ha-Nhat Cuong, Van-Dang Tran, Linh~Ngo Van, and Khoat Than.
\newblock Eliminating overfitting of probabilistic topic models on short and
  noisy text: The role of dropout.
\newblock {\em International Journal of Approximate Reasoning}, 2019.

\bibitem{Dillahunt:2017}
Tawanna~R. Dillahunt, Xinyi Wang, Earnest Wheeler, Hao~Fei Cheng, Brent Hecht,
  and Haiyi Zhu.
\newblock The sharing economy in computing: A systematic literature review.
\newblock {\em Proc. ACM Hum.-Comput. Interact.}, 1(CSCW):38:1--38:26, December
  2017.

\bibitem{fang2016effect}
Bin Fang, Qiang Ye, and Rob Law.
\newblock Effect of sharing economy on tourism industry employment.
\newblock {\em Annals of Tourism Research}, 57:264--267, 2016.

\bibitem{fleiss1971measuring}
Joseph~L Fleiss.
\newblock Measuring nominal scale agreement among many raters.
\newblock {\em Psychological bulletin}, 76(5):378, 1971.

\bibitem{fradkin2015bias}
Andrey Fradkin, Elena Grewal, Dave Holtz, and Matthew Pearson.
\newblock Bias and reciprocity in online reviews: Evidence from field
  experiments on airbnb.
\newblock In {\em Proc. of EC '15}, pages 641--641. ACM, 2015.

\bibitem{fradkin2017determinants}
Andrey Fradkin, Elena Grewal, and David Holtz.
\newblock The determinants of online review informativeness: Evidence from
  field experiments on airbnb.
\newblock Technical report, Working Paper, 2017.

\bibitem{goldberg2014word2vec}
Yoav Goldberg and Omer Levy.
\newblock word2vec explained: deriving mikolov et al.'s negative-sampling
  word-embedding method.
\newblock {\em arXiv preprint arXiv:1402.3722}, 2014.

\bibitem{guttentag2018tourists}
Daniel Guttentag, Stephen Smith, Luke Potwarka, and Mark Havitz.
\newblock Why tourists choose airbnb: A motivation-based segmentation study.
\newblock {\em Journal of Travel Research}, 57(3):342--359, 2018.

\bibitem{hage1980theories}
Jerald Hage.
\newblock {\em Theories of organizations: Form, process, and transformation}.
\newblock Wiley New York, 1980.

\bibitem{haukoos2005advanced}
Jason~S Haukoos and Roger~J Lewis.
\newblock Advanced statistics: bootstrapping confidence intervals for
  statistics with ``difficult'' distributions.
\newblock {\em Academic emergency medicine}, 12(4):360--365, 2005.

\bibitem{heo2019happening}
Cindy~Yoonjoung Heo, In{\`e}s Blal, and Miju Choi.
\newblock What is happening in paris? airbnb, hotels, and the parisian market:
  a case study.
\newblock {\em Tourism Management}, 70:78--88, 2019.

\bibitem{ikkala2015monetizing}
Tapio Ikkala and Airi Lampinen.
\newblock Monetizing network hospitality: Hospitality and sociability in the
  context of airbnb.
\newblock In {\em Proc. of CSCW '15}, pages 1033--1044. ACM, 2015.

\bibitem{jelodar2019latent}
Hamed Jelodar, Yongli Wang, Chi Yuan, Xia Feng, Xiahui Jiang, Yanchao Li, and
  Liang Zhao.
\newblock Latent dirichlet allocation (lda) and topic modeling: models,
  applications, a survey.
\newblock {\em Multimedia Tools and Applications}, 78(11):15169--15211, 2019.

\bibitem{joseph2019analyzing}
George Joseph and Vinu Varghese.
\newblock Analyzing airbnb customer experience feedback using text mining.
\newblock In {\em Big Data and Innovation in Tourism, Travel, and Hospitality},
  pages 147--162. Springer, 2019.

\bibitem{jung2016social}
Jiwon Jung, Susik Yoon, SeungHyun Kim, Sangkeun Park, Kun-Pyo Lee, and Uichin
  Lee.
\newblock Social or financial goals?: Comparative analysis of user behaviors in
  couchsurfing and airbnb.
\newblock In {\em Proc. of CHI '16}, pages 2857--2863. ACM, 2016.

\bibitem{ketchen1996application}
David~J Ketchen and Christopher~L Shook.
\newblock The application of cluster analysis in strategic management research:
  an analysis and critique.
\newblock {\em Strategic management journal}, 17(6):441--458, 1996.

\bibitem{klein2017quality}
Maximilian Klein, Jinhao Zhao, Jiajun Ni, Isaac Johnson, Benjamin~Mako Hill,
  and Haiyi Zhu.
\newblock Quality standards, service orientation, and power in airbnb and
  couchsurfing.
\newblock {\em Proceedings of the ACM on Human-Computer Interaction},
  1(CSCW):58, 2017.

\bibitem{lampinen2016hosting}
Airi Lampinen and Coye Cheshire.
\newblock Hosting via airbnb: Motivations and financial assurances in monetized
  network hospitality.
\newblock In {\em Proc. of CHI '16}, pages 1669--1680. ACM, 2016.

\bibitem{lawani2018reviews}
Abdelaziz Lawani, Michael Michael~R Reed, Tyler Mark, and Yuqing Zheng.
\newblock Reviews and price on online platforms: Evidence from sentiment
  analysis of airbnb reviews in boston.
\newblock {\em Regional Science and Urban Economics}, 2018.

\bibitem{lee2019analysing}
Carmen Kar~Hang Lee, Ying~Kei Tse, Minhao Zhang, and Jie Ma.
\newblock Analysing online reviews to investigate customer behaviour in the
  sharing economy.
\newblock {\em Information Technology \& People}, 2019.

\bibitem{lin2019spend}
Pearl~MC Lin, Daisy~XF Fan, Hanqin~Qiu Zhang, and Chloe Lau.
\newblock Spend less and experience more: Understanding tourists’ social
  contact in the airbnb context.
\newblock {\em International Journal of Hospitality Management}, 83:65--73,
  2019.

\bibitem{liu2018impact}
Jianwei Liu, Karen Xie, Qiang Ye, and Dong Jing.
\newblock The impact of sharing economy on local employment: Evidence from
  airbnb.
\newblock 2018.

\bibitem{luo2018airbnb}
Yi~Luo.
\newblock What airbnb reviews can tell us? an advanced latent aspect rating
  analysis approach.
\newblock 2018.

\bibitem{luo2019understanding}
Yi~Luo and Rebecca~Liang Tang.
\newblock Understanding hidden dimensions in textual reviews on airbnb: An
  application of modified latent aspect rating analysis (lara).
\newblock {\em International Journal of Hospitality Management}, 80:144--154,
  2019.

\bibitem{martin2018modelling}
Eva Martin-Fuentes, Cesar Fernandez, Carles Mateu, and Estela Marine-Roig.
\newblock Modelling a grading scheme for peer-to-peer accommodation: Stars for
  airbnb.
\newblock {\em International Journal of Hospitality Management}, 69:75--83,
  2018.

\bibitem{moore1999crossing}
Geoffrey~A Moore and Regis McKenna.
\newblock Crossing the chasm.
\newblock 1999.

\bibitem{pennebaker2001linguistic}
James~W Pennebaker, Martha~E Francis, and Roger~J Booth.
\newblock Linguistic inquiry and word count: Liwc 2001.
\newblock {\em Mahway: Lawrence Erlbaum Associates}, 71(2001):2001, 2001.

\bibitem{quattrone2016airbnb}
Giovanni Quattrone, Davide Proserpio, Daniele Quercia, Licia Capra, and Mirco
  Musolesi.
\newblock Who benefits from the sharing economy of airbnb?
\newblock In {\em Proc. of WWW '16}, pages 1385--1394, 2016.

\bibitem{rogers2010diffusion}
Everett~M Rogers.
\newblock {\em Diffusion of innovations}.
\newblock Simon and Schuster, 2010.

\bibitem{roma2019sharing}
Paolo Roma, Umberto Panniello, and Giovanna~Lo Nigro.
\newblock Sharing economy and incumbents' pricing strategy: The impact of
  airbnb on the hospitality industry.
\newblock {\em International Journal of Production Economics}, 214:17--29,
  2019.

\bibitem{satama2014consumer}
Sampo Satama et~al.
\newblock Consumer adoption of access-based consumption services-case airbnb.
\newblock 2014.

\bibitem{shabrina2019airbnb}
Zahratu Shabrina, Elsa Arcaute, and Michael Batty.
\newblock Airbnb's disruption of the housing structure in london.
\newblock {\em arXiv preprint arXiv:1903.11205}, 2019.

\bibitem{tausczik2010psychological}
Yla~R Tausczik and James~W Pennebaker.
\newblock The psychological meaning of words: Liwc and computerized text
  analysis methods.
\newblock {\em Journal of language and social psychology}, 29(1):24--54, 2010.

\bibitem{van2000research}
Andrew~H Van~de Ven, Harold~L Angle, and Marshall~Scott Poole.
\newblock {\em Research on the management of innovation: The Minnesota
  studies}.
\newblock Oxford University Press on Demand, 2000.

\bibitem{wachsmuth2018airbnb}
David Wachsmuth and Alexander Weisler.
\newblock Airbnb and the rent gap: Gentrification through the sharing economy.
\newblock {\em Environment and Planning A: Economy and Space},
  50(6):1147--1170, 2018.

\bibitem{zaltman1973innovations}
Gerald Zaltman, Robert Duncan, and Jonny Holbek.
\newblock {\em Innovations and organizations}.
\newblock John Wiley \& Sons, 1973.

\bibitem{zervas2015first}
Georgios Zervas, Davide Proserpio, and John Byers.
\newblock A first look at online reputation on airbnb, where every stay is
  above average.
\newblock 2015.

\bibitem{Zervas2016rise}
Georgios Zervas, Davide Proserpio, and John Byers.
\newblock The rise of the sharing economy: Estimating the impact of airbnb on
  the hotel industry.
\newblock {\em Boston U. School of Management Research Paper}, (2013-16), 2016.

\bibitem{zervas2015impact}
Georgios Zervas, Davide Proserpio, and John~W Byers.
\newblock The impact of the sharing economy on the hotel industry: Evidence
  from airbnb's entry into the texas market.
\newblock In {\em Proc. of EC '15}, pages 637--637. ACM, 2015.

\end{thebibliography}

\end{document}